\documentclass[twocolumn]{aastex631}
\shorttitle{[C\,{\sc ii}] of TN J0924$-$2201}
\shortauthors{Lee et al.}
\begin{document}

\title{Ongoing and fossil large-scale outflows detected in a high-redshift radio galaxy: [C\,II] observations of TN J0924$-$2201 at $z=5.174$}

\correspondingauthor{Kianhong Lee}
\email{kianhong.lee@astr.tohoku.ac.jp}

\author[0000-0003-4814-0101]{Kianhong Lee}
\affiliation{Astronomical Institute, Tohoku University, Aramaki, Aoba-ku, Sendai, Miyagi 980-8578, Japan}
\affiliation{National Astronomical Observatory of Japan, 2-21-1 Osawa, Mitaka, Tokyo 181-8588, Japan}
\affiliation{Institute of Astronomy, Graduate School of Science, The University of Tokyo, 2-21-1 Osawa, Mitaka, Tokyo 181-0015, Japan}

\author[0000-0003-4814-0101]{Masayuki Akiyama}
\affiliation{Astronomical Institute, Tohoku University, Aramaki, Aoba-ku, Sendai, Miyagi 980-8578, Japan}

\author[0000-0002-4052-2394]{Kotaro Kohno}
\affiliation{Institute of Astronomy, Graduate School of Science, The University of Tokyo, 2-21-1 Osawa, Mitaka, Tokyo 181-0015, Japan}
\affiliation{Research Center for the Early Universe, School of Science, The University of Tokyo, 7-3-1 Hongo, Bunkyo-ku, Tokyo 113-0033, Japan}

\author[0000-0002-2364-0823]{Daisuke Iono}
\affiliation{National Astronomical Observatory of Japan, 2-21-1 Osawa, Mitaka, Tokyo 181-8588, Japan}
\affiliation{Department of Astronomy, School of Science, The Graduate University for Advanced Studies (SOKENDAI), Osawa, Mitaka, Tokyo 181-8588, Japan}

\author[0000-0001-6186-8792]{Masatoshi Imanishi}
\affiliation{National Astronomical Observatory of Japan, 2-21-1 Osawa, Mitaka, Tokyo 181-8588, Japan}
\affiliation{Department of Astronomy, School of Science, The Graduate University for Advanced Studies (SOKENDAI), Osawa, Mitaka, Tokyo 181-8588, Japan}

\author[0000-0001-6469-8725]{Bunyo Hatsukade}
\affiliation{National Astronomical Observatory of Japan, 2-21-1 Osawa, Mitaka, Tokyo 181-8588, Japan}
\affiliation{Institute of Astronomy, Graduate School of Science, The University of Tokyo, 2-21-1 Osawa, Mitaka, Tokyo 181-0015, Japan}

\author[0000-0003-1937-0573]{Hideki Umehata}
\affiliation{Institute for Advanced Research, Nagoya University, Furocho, Chikusa, Nagoya 464-8602, Japan}
\affiliation{Department of Physics, Graduate School of Science, Nagoya University, Furocho, Chikusa,
Nagoya 464-8602, Japan}

\author[0000-0002-7402-5441]{Tohru Nagao}
\affiliation{Research Center for Space and Cosmic Evolution, Ehime University, 2-5 Bunkyo-cho, Matsuyama, Ehime 790-8577, Japan}

\author[0000-0002-3531-7863]{Yoshiki Toba}
\affiliation{National Astronomical Observatory of Japan, 2-21-1 Osawa, Mitaka, Tokyo 181-8588, Japan}
\affiliation{Academia Sinica Institute of Astronomy and Astrophysics, 11F of Astronomy-Mathematics Building, AS/NTU, No.1, Section 4, Roosevelt Road, Taipei 10617, Taiwan}
\affiliation{Research Center for Space and Cosmic Evolution, Ehime University, 2-5 Bunkyo-cho, Matsuyama, Ehime 790-8577, Japan}

\author[0000-0003-2682-473X]{Xiaoyang Chen}
\affiliation{National Astronomical Observatory of Japan, 2-21-1 Osawa, Mitaka, Tokyo 181-8588, Japan}

\author[0000-0002-1639-1515]{Fumi Egusa}
\affiliation{Institute of Astronomy, Graduate School of Science, The University of Tokyo, 2-21-1 Osawa, Mitaka, Tokyo 181-0015, Japan}

\author[0000-0002-4377-903X]{Kohei Ichikawa}
\affiliation{Faculty of Science and Engineering, Waseda University, 3-4-1 Okubo, Shinjuku, Tokyo 169-8555, Japan}
\affiliation{Astronomical Institute, Tohoku University, Aramaki, Aoba-ku, Sendai, Miyagi 980-8578, Japan}
\affiliation{Frontier Research Institute for Interdisciplinary Sciences, Tohoku University, Sendai 980-8578, Japan}

\author[0000-0001-9452-0813]{Takuma Izumi}
\affiliation{National Astronomical Observatory of Japan, 2-21-1 Osawa, Mitaka, Tokyo 181-8588, Japan}

\author[0000-0002-8299-0006]{Naoki Matsumoto}
\affiliation{Astronomical Institute, Tohoku University, Aramaki, Aoba-ku, Sendai, Miyagi 980-8578, Japan}

\author[0000-0001-7825-0075]{Malte Schramm}
\affiliation{Universit$\ddot{a}$t Potsdam, Karl-Liebknecht-Str. 24/25, D-14476 Potsdam, Germany}
%\affiliation{Graduate School of Science and Engineering, Saitama University, Shimo-Okubo 255, Sakura-ku, Saitama-shi, Saitama 338-8570, Japan}

\author[0000-0002-2689-4634]{Kenta Matsuoka}
\affiliation{Utena Meishu Company, Limited, 384 Yanagihara, Matsuyama, Ehime 799-2434, Japan}



\begin{abstract}

We present Atacama Large Millimeter/submillimeter Array observations of the [C\,{\sc ii}]\,158\,$\mu$m line and the underlying continuum emission of TN J0924$-$2201, which is one of the most distant known radio galaxies at $z>5$.
The [C\,{\sc ii}] line and 1-mm continuum emission are detected at the host galaxy.  
The systemic redshift derived from the [C\,{\sc ii}] line is $z_{\rm [C\,{\sc II}]}=5.1736\pm0.0002$, 
indicating that the Ly$\alpha$ line is redshifted by a velocity of $1035\pm10$\,km\,s$^{-1}$, marking the largest velocity offset between the [C\,{\sc ii}] and Ly$\alpha$ lines recorded at $z>5$ to date. 
In the central region of the host galaxy, we identified a redshifted substructure of [C\,{\sc ii}] with a velocity of $702\pm17$\,km\,s$^{-1}$, which is close to the C\,{\sc iv} line with a velocity of $500\pm10$\,km\,s$^{-1}$. The position and the velocity offsets align with a model of an outflowing shell structure, consistent with the large velocity offset of Ly$\alpha$. 
The non-detection of [C\,{\sc ii}] and dust emission from the three CO(1--0)-detected companions indicates their different nature compared to dwarf galaxies based on the photodissociation region model.
Given their large velocity of $\sim1500$\,km\,s$^{-1}$, outflowing molecular clouds induced by the AGN is the most plausible interpretation, and they may exceed the escape velocity of a $10^{13}\,M_{\odot}$ halo.
These results suggest that TN J0924$-$2201, with the ongoing and fossil large-scale outflows, is in a distinctive phase of removing molecular gas from a central massive galaxy in an overdense region in the early universe.
A dusty H\,{\sc i} absorber at the host galaxy is an alternative interpretation.

\end{abstract}



\section{Introduction} \label{sec:intro}

High-redshift ($z\gtrsim2$) radio galaxies are massive galaxies \citep[stellar mass $M_{*}>10^{11}\,M_{\odot}$;][]{DeBreuck10}, containing powerful radio-loud (radio luminosity at rest-frame 500 MHz $L_{\rm 500 MHz}>10^{27}\,\rm W\,Hz^{-1}$) active galactic nuclei (AGNs). Since high-$z$ radio galaxies are often found in overdense regions, they have been used as beacons for searching protoclusters \citep{MileyDeBeruck08}. 
Their evolution appears to be related with the evolution of the overdense regions, or the protoclusters \citep{Magliocchetti22}.
Regarding their stellar properties, high-$z$ radio galaxies are still star-forming but found to be on the way to be quenched \citep{Falkendal19}. 
The spatial distribution of high-$z$ radio galaxies and molecular gas shows intriguing phenomenons.
CO emission lines have been detected with spatial offsets from the radio galaxies, and the preferential alignments between CO emission and radio jet axis have been discovered at $z\sim$2--5 \citep{Klamer04,Klamer05,Emonts14,Lee23}.
Several interpretations such as jet-induced metal enrichment, merger, outflow \citep{Klamer04,Emonts14}, and observational evidences such as inflow \citep{Emonts23a} and gas-induced synchrotron brightening \citep{Emonts23b} have been suggested as the physical origins of the phenomenons, based on the observations of radio galaxies at $z\sim$2--4.

At $z\sim5$, TN J0924$-$2201 is one of the most distant known radio galaxies, and it has been discovered and spectroscopically identified for more than two decades \citep[$z_{\rm Ly\alpha}=5.20$;][]{vanBreugel99}. 
The detection of Ly$\alpha$ line ($z_{\rm Ly\alpha}=5.195$) and C\,{\sc iv} 1549{\,\AA} emission line ($z_{\rm C\,{\sc IV}}=5.184$) from this galaxy has also been reported by \cite{Matsuoka11}.
TN J0924$-$2201 has massive host galaxy with stellar mass $M_{*}=10^{11.10}\,M_{\odot}$, which is estimated with a spectral energy distribution (SED) fitting to multi-band photometry including mid-infrared bands \citep{DeBreuck10}.
In addition, TN J0924$-$2201 is located at an overdense region of Ly$\alpha$ emitters \citep{Venemans04} and Lyman-break galaxies \citep{Overzier06}, extending roughly $\sim3'$ which corresponds to physical size of $\rm \sim1\,Mpc$.
Regarding the molecular gas properties, using the Australia Telescope Compact Array (ATCA), \cite{Klamer05} have reported the detection of CO(1--0) and CO(5--4) lines, whose luminosity indicates the existence of a massive ($M_{\rm H_{2}}\sim10^{11}\,M_{\odot}$) molecular gas reservoir with tentative spatial offset ($\sim3''$--$5''$) from the host galaxy.
Almost two decades later, using the Karl G. Jansky Very Large Array (VLA), \cite{Lee23} have confirmed the spatial offset ($\sim2''$--$5''$, corresponding to physical 12--33 kpc), and further spatially resolved the CO(1--0) line emission into three components. 
Each component has molecular mass of (2--4)$\times10^{10}(\frac{\alpha_{\rm CO}}{0.8})\,M_{\odot}$, where $\alpha_{\rm CO}$ is the CO-to-H$_{2}$ conversion factor. The nearest one from the host galaxy appears to spatially align with the radio jet axis.
Such large spatial offset between the host galaxy of TN J0924$-$2201 and the three molecular gas companions, together with the jet-gas alignment, showing a typical spatial distribution like other radio galaxies at lower-$z$ \citep{Lee23}. 
However, the relation between TN J0924$-$2201 and the three CO(1--0) companions is still inconclusive due to the lack of the kinetic connection between the host galaxy and the companions.

To investigate the nature of TN J0924$-$2201, it is important to obtain the kinematic properties of the interstellar medium (ISM) in both the host galaxy and the companions.
[C\,{\sc ii}] 158 $\mu$m fine structure line has been known as the best kinematic tracer of the global ISM properties in high-redshift galaxies thanks to its brightness and moderate excitation condition. Together with other emission lines and infrared (IR) luminosity, [C\,{\sc ii}] line can provide a diagnostic of the neutral gas, HII regions, and photodissociation regions (PDRs) in galaxies \citep[e.g.,][]{Casey14}.
Therefore, targeting on [C\,{\sc ii}] 158 $\mu$m line and the underlying observed-frame far-infrared (FIR) 1-mm continuum emission, we conducted the observation of TN J0924$-$2201 with the Atacama Large Millimeter/submillimeter Array (ALMA). 

In this paper, we describe the observations in Section~\ref{sec:obs}, and the results in Section~\ref{sec:results}. Based on the results, we then discuss the interpretations in Section~\ref{sec:discussion}, and conclude in Section~\ref{sec:conclusions}.
Throughout this paper, we adopt the AB magnitude system and
assume a standard $\Lambda$CDM cosmology with $H_{0}=70\,\rm km\,s^{-1}\,Mpc^{-1}$, $\Omega_{\rm M}=0.3$ and $\Omega_{\rm \Lambda}=0.7$. 
At $z=5.2$, 1$''$ corresponds to the physical scale of $\sim$6.2 kpc.

\section{ALMA Observations} \label{sec:obs}

Targeting on the redshifted [C\,{\sc ii}] 158 $\mu$m fine structure line and the underlying observed-frame 1-mm continuum emission of the radio galaxy, TN J0924$-$2201,
the ALMA Band-7 observations (ID: 2021.1.00219.S; Principal Investigator: K. Lee) were conducted on 2022 Jan 02 and 03.
The primary calibrators are J1037$-$2934 and J1058$+$0133.
The phase calibrator is J0927$-$2034.
The on-source integration time is 1.8 hours. 
The total bandwidth 7.5 GHz consists of four spectral windows of 1.875 GHz. Each spectral window consists of 240 channels, and the bandwidth of each channel is 7812.5 kHz.
The lower sideband covers 293.59--297.23 GHz, 
and the upper sideband covers 305.41--309.11 GHz, corresponding to wavelength $\sim1\,$mm. 

Data reduction was performed with Common Astronomy Software Applications \cite[CASA;][]{CASA22} version 6.5.1 in the standard manner. 
To avoid the contamination of line emission, the continuum image was produced with line-emission-free spectral windows and channels (293.59--306.62 GHz). We applied {\tt tclean} with deconvolver {\tt mtmfs} and Briggs weighting {\tt robust=0.5}. The pixel scale is set to $0''.1$.
The synthesized beam size is 0$''$.38 $\times$ 0$''$.28 with Position Angle (P.A.) = $-$78$^{\circ}$.4. The rms noise level is 19 $\rm \mu Jy\,beam^{-1}$.

To obtain the spectral data cube, firstly we used CASA task {\tt uvcontsub} to subtract the continuum. We used two spectral windows in the lower sideband and about a half of one spectral window in the upper sideband for continuum fitting with {\tt fitorder=1}, because the line emission mostly appears in the other spectral window in the upper sideband. After subtracting the continuum, we applied {\tt tclean} with deconvolver {\tt hogbom} and Briggs weighting {\tt robust=2.0}. The pixel scale is set to $0''.1$.
The synthesized beam size is 0$''$.48 $\times$ 0$''$.36 with P.A. = $-$79$^{\circ}$.5. 
By combining different numbers of channels, we made data cubes in two velocity resolutions, 91.2 and 60.8 $\rm km\,s^{-1}$, and we use them for discussion on detection and line profile in the later sections, respectively.
The resulting rms noise level is 
74.5 and 91.2 $\rm \mu Jy\,beam^{-1}\,channel^{-1}$, where one channel corresponds to 91.2 and 60.8 $\rm km\,s^{-1}$, respectively.

\section{Results} \label{sec:results}

\subsection{1-mm continuum emission}

\begin{figure*}[ht!]
\epsscale{1.2}
\plotone{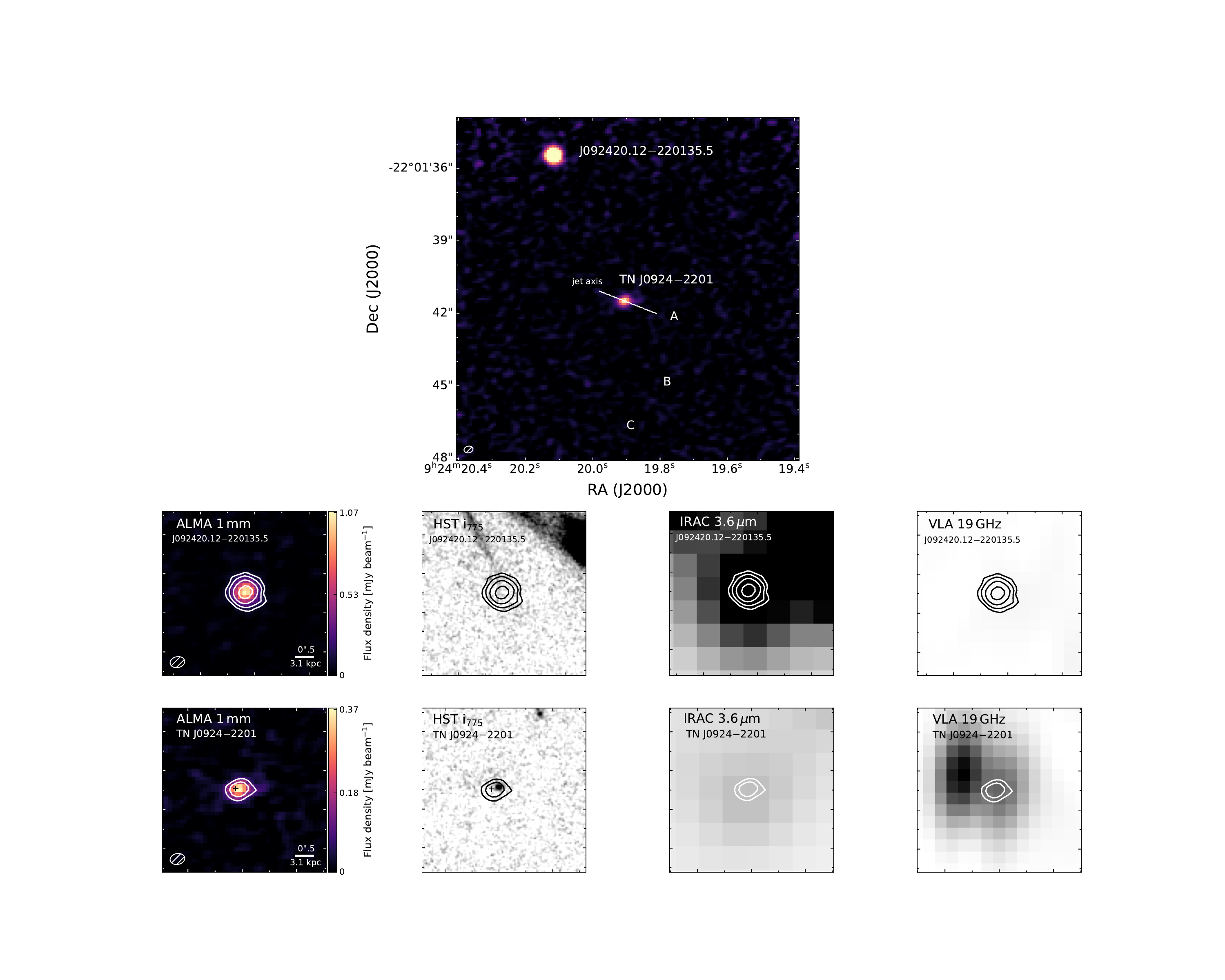}
\caption{\textbf{Top}: ALMA 1$\,$mm continuum map. TN J0924$-$2201 is at the center of the image and the source at the north is J092420.12$-$220135.5. White letters, ``A'', ``B'' and ``C'' indicate the positions of CO(1--0) detection. 
\textbf{Bottom}: The zoomed out continuum images of J092420.12$-$220135.5 and TN J0924$-$2201. The white and black contours indicate 5$\sigma$, 10$\sigma$, 20$\sigma$ and 40$\sigma$, where 1$\sigma=19$ $\mu$m$\,$beam$^{-1}$. Gray-scaled HST/ACS F775W ($i$-band), IRAC 3.6 $\mu$m and VLA 19 GHz images of J092420.12$-$220135.5 and TN J0924$-$2201 are shown. The black crosses in the lower ALMA and HST images indicate the Gaia-corrected optical peak of TN J0924$-$2201.
The ellipses in the bottom left corner of ALMA images indicate the synthesized beams.
\label{fig:cont}}
\end{figure*}

\begin{figure}[ht!]
\epsscale{1}
\plotone{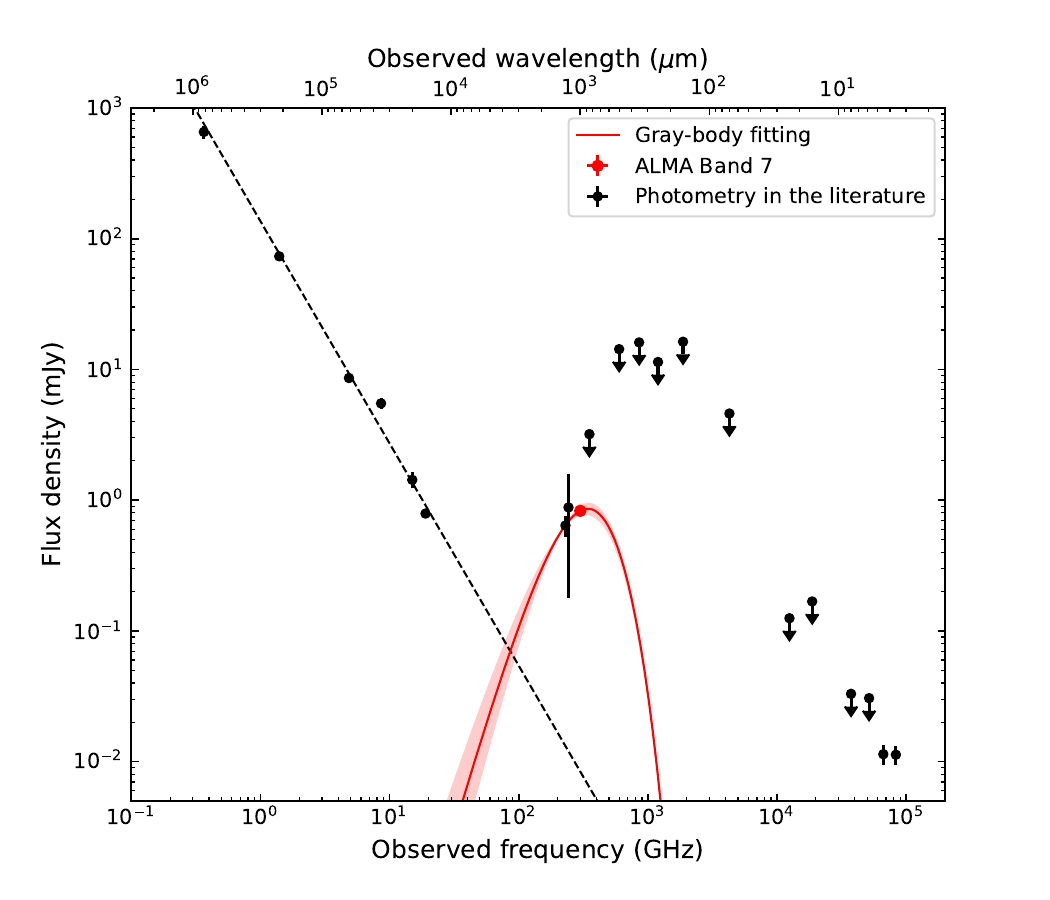}
\caption{The spectral energy distribution (SED) of TN J0924$-$2201.
Black dots indicate photometries (or $3\sigma$ upper limits with down arrows) in the literature \citep[][and references therein]{Lee23}. The red dot indicates ALMA Band-7 data. The red solid line indicates the best fit of modified blackbody fitting, and the red shaded area indicates 1$\sigma$ uncertainty. The black dashed line indicates the power law of non-thermal synchrotron emission (spectral index $\alpha=-1.7$).
\label{fig:SED}}
\end{figure}

Figure~\ref{fig:cont} shows the ALMA 1-mm continuum map. The map is centered at the position of the detection of TN J0924$-$2201, while a brighter source, J092420.12$-$220135.5, is also detected at the north. This object is a serendipitous discovery reported by \cite{Lee23} based on the detection in ALMA 1.3 mm continuum. 
The peak flux of TN J0924$-$2201 is $392\pm19\,\mu$Jy and the integrated flux is $829\pm57\,\mu$Jy. 
The peak flux of J092420.12$-$220135.5 is $1.27\pm0.03\,$mJy and the integrated flux is $2.61\pm0.10\,$mJy.
In Figure~\ref{fig:cont}, we marked the positions where CO(1--0) line is detected \citep[VLA 19-GHz observations;][]{Lee23} in letters A, B and C. We found no 1-mm continuum emission counterparts at these positions.
The $3\sigma$ upper limit of non-detection is $57\,\mu$Jy with beam $\sim0''.4$.
We are aware that there are two archival ALMA Band-6 ($\sim1.3\,$mm) snapshots of TN J0924$-$2201 with integration time of two and eight minutes. 
No 1.3-mm continuum emission is detected at the positions of A, B, and C.

TN J0924$-$2201 is detected in all of the HST/ACS F775W ($i$-band), IRAC 3.6 $\mu$m and VLA 19 GHz images. 
By comparing the position of a nearby star in Gaia catalog \citep{Gaia21}, 
\cite{Lee23} have reported that {\it HST} image has astrometric offset (RA: $-0''.17$, Dec: $+0''.06$). 
We indicate the astrometrically corrected position of the optical peak in the figures including Figure~\ref{fig:cont}.

For J092420.12$-$220135.5, it is detected in neither $i$-band nor 19 GHz. Considering the redshift dependence of the flux ratio of $S_{\rm 3.6\,\mu m}/S_{\rm 1.2\, mm}$ of submillimeter galaxies (SMGs) \citep{Yamaguchi19}, this {\it HST}-dark SMG is likely at $z<4$. Therefore, this object is considered not to be associated with the overdense region surrounding TN J0924$-$2201.

Figure \ref{fig:SED} shows the spectral energy distribution (SED) of TN J0924$-$2201. Photometries are from IRAC, IRS, MIPS \citep{DeBreuck10}, PACS, SPIRE \citep{Drouart14}, SCUBA \citep{Reuland04}, WSRT, VLA \citep{DeBreuck00,Falkendal19,Lee23}, ALMA Band 6 \citep{Falkendal19,Lee23} and ALMA Band 7 (this work).
Based on the fluxes at 365 MHz and 19 GHz, the spectral index of non-thermal synchrotron emission can be derived as $\alpha=-1.7$ (assuming $S_{\nu}\propto\nu^{\alpha}$).
Given the extrapolation of the power law,
the contribution of synchrotron emission is negligible at observed 1$\,$mm \citep{Lee23}.
Besides, the contribution of AGN-heating dust generally peaks at rest-frame 5--35$\,\mu$m, corresponding to observed $\sim$30--220$\,\mu$m for TN J0924$-$2201, and the flux drops dramatically with increasing wavelength \citep{Shi14}.
Therefore, the flux at observed 1$\,$mm is dominated by the thermal dust emission from the star formation activity.

With ALMA Band-6 (1.3$\,$mm) and Band-7 (1$\,$mm) data, we applied modified blackbody fitting on TN J0924$-$2201. 
Following \cite{Casey12} and \cite{Falkendal19}, the modified blackbody (i.e., a ``graybody'') is:
\begin{equation}
    S_{\rm BB}=N_{\rm BB}(1-e^{-(\nu/\nu_{0})^{\beta}})\frac{2\pi h}{c^{2}}\frac{\nu^{3}}{(e^{h\nu/k_{\rm B}T_{\rm dust}}-1)},
\end{equation}
where $N_{\rm BB}$ is the normalization number, $\nu$ is the rest-frame frequency, $\nu_{0}$ is the critical frequency where the source becomes optically thin, being assumed to be 1.5 THz \citep{Conley11}, $\beta$ is the emissivity, and $T_{\rm dust}$ is the dust temperature.

The best fit is shown in red solid line with $1\sigma$ uncertainty indicated in red shaded area (Figure~\ref{fig:SED}). 
We obtained $T_{\rm dust}=29.3\pm0.4\,\rm K$ and $\beta=1.51\pm0.46$.
This $T_{\rm dust}$ is consistent with the mean value of local luminous and ultra-luminous infrared galaxies (ULIRGs) \citep{U12}.
We then derived the 8--1000$\,\mu$m infrared luminosity $L_{\rm IR}=1.72^{+0.13}_{-0.14}\times10^{12}\,L_{\odot}$, 
which indicates TN J0924$-$2201 is an ULIRG, consistent with the estimation of \cite{Falkendal19} and \cite{Lee23}.

\subsection{[C\,{\sc ii}] line emission} \label{sec:cii}

\begin{figure*}[ht!]
\epsscale{1.1}
\plotone{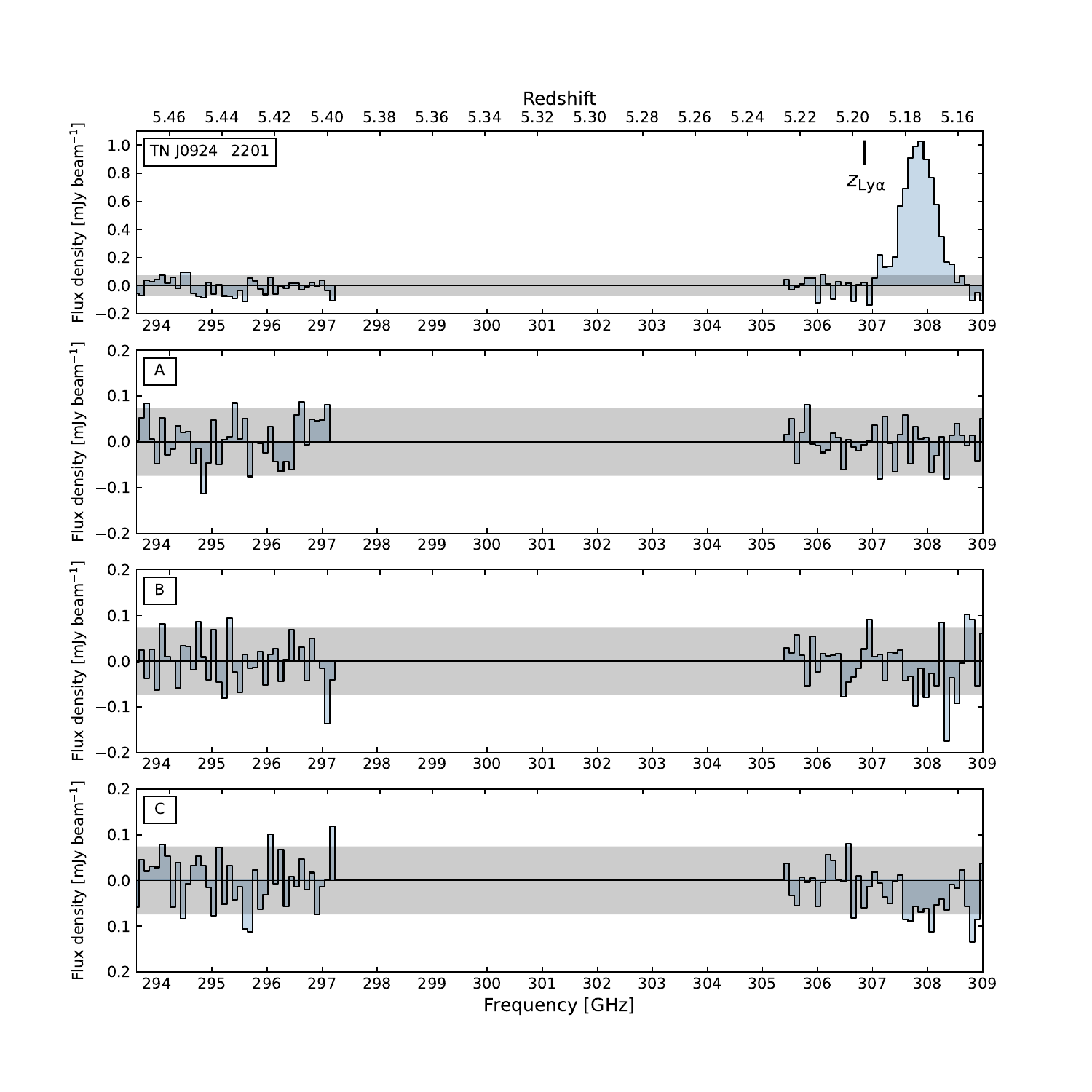}
\caption{Spectra of TN J0924$-$2201 and three CO(1--0) detected regions, A, B and C, which are indicated in Figure ~\ref{fig:cont}, with channel width $\sim$91.2 km$\,$s$^{-1}$. 
The shaded areas indicate the $1\,\sigma$ noise level.
In the panel of TN J0924$-$2201, the offset of redshift between the peak of [C\,{\sc ii}] line and Ly$\alpha$ is indicated.
\label{fig:4spectra}}
\end{figure*}

\begin{figure*}[ht!]
\epsscale{1.25}
\plotone{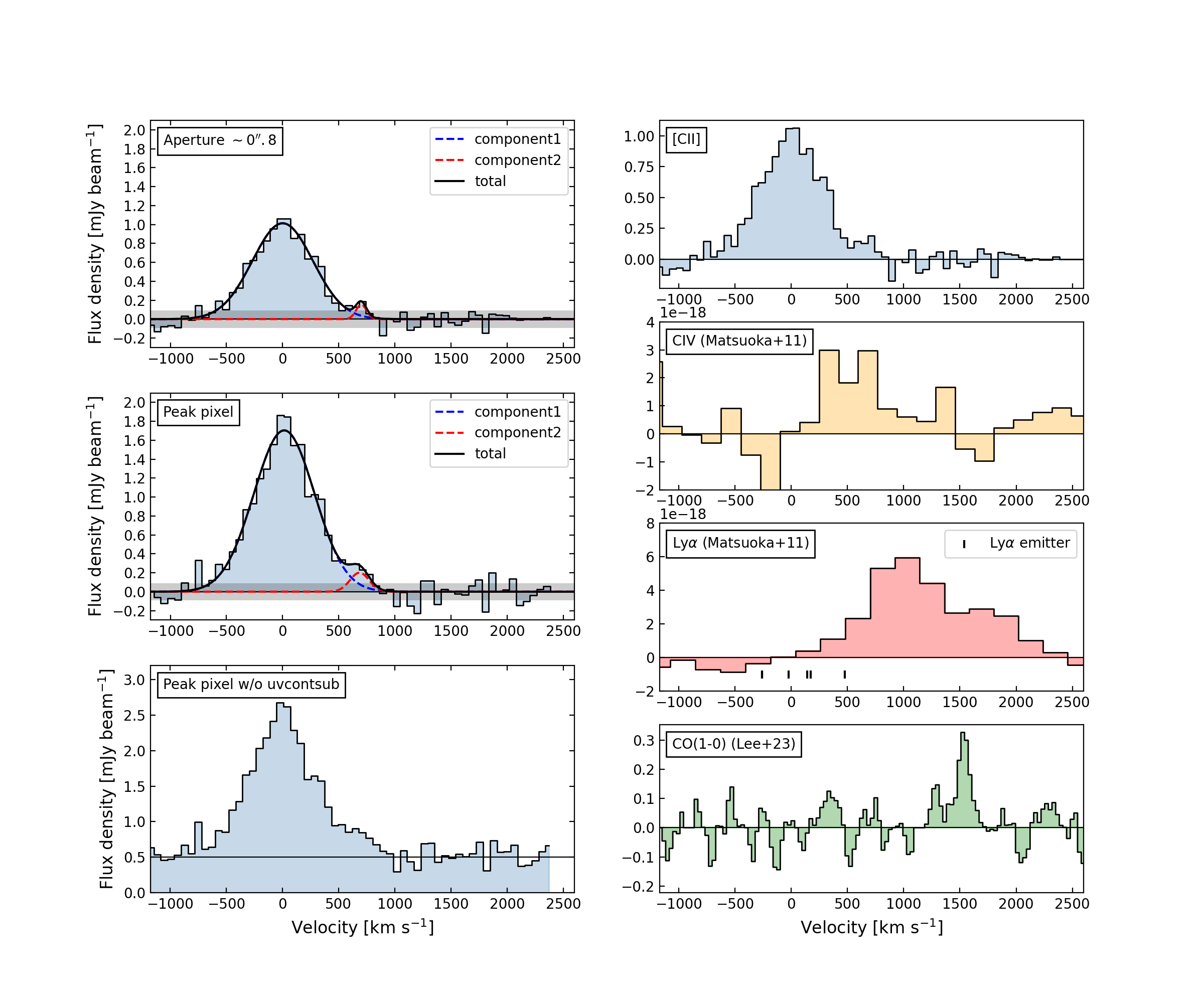}
\caption{
\textbf{Left}:
Spectra of TN J0924$-$2201 with channel width $\sim$60.8 km$\,$s$^{-1}$.
The upper left panel shows the spectrum extracted with aperture $\sim0''.8$ and the middle left panel shows the spectrum at the peak pixel, whose size is $0''.1\times0''.1$. 
The gray shaded areas indicate the $1\,\sigma$ noise level. 
Both spectra are fitted with two Gaussian components. Component 1 and component 2 are shown with blue and red dashed lines, respectively. The sum of the two components is indicated as ``total'' in black solid line.
The peak of component 1 is defined as the systemic velocity 0 km$\,$s$^{-1}$.
The bottom left panel shows the spectrum at the peak pixel in the spectral cube without the subtraction of the continuum and the contamination from the nearby bright source. The horizontal line at flux density of 0.5\,mJy\,beam$^{-1}$ roughly indicates the continuum level.
\textbf{Right}:
Four spectra of [C\,{\sc ii}], C\,{\sc iv}, Ly$\alpha$ and CO(1--0) are shown in the right panels from top to bottom.
The unit of vertical axis of [C\,{\sc ii}] and CO(1--0) is flux density in mJy$\,$beam$^{-1}$, and the unit of vertical axis of C\,{\sc iv} and Ly$\alpha$ is flux density in $\rm erg\,s^{-1}\,cm^{-2}\AA^{-1}$.
In the panel of Ly$\alpha$, the velocities of six Ly$\alpha$ emitters in the overdense region around TN J0924$-$2201 reported by \cite{Venemans04} are indicated in black vertical bars.
C\,{\sc iv} and Ly$\alpha$ spectra were obtained with Subaru/FOCAS by \citet{Matsuoka11}.
CO(1--0) spectrum with beam size $\sim4''.4$ was obtained with VLA by \citet{Lee23}.
\label{fig:linefit}}
\end{figure*}

\begin{figure*}[ht!]
\epsscale{1.2}
\plotone{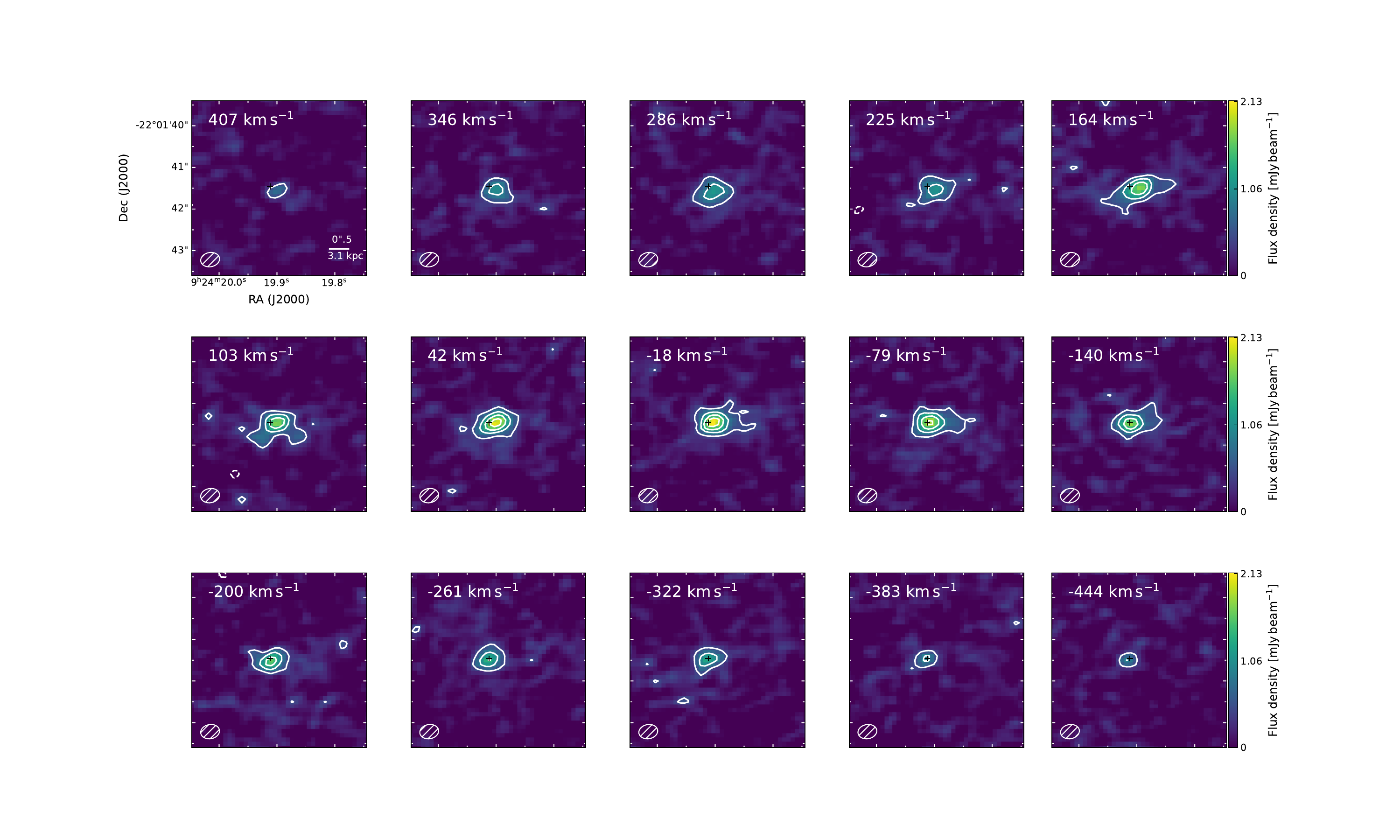}
\caption{Channal map of [C\,{\sc ii}]. Velocities are indicated in the top left corner of each image in white texts. White contours indicate $5\,\sigma$, $10\,\sigma$, $15\,\sigma$ and $20\,\sigma$, where $1\,\sigma=0.091\,\rm mJy\,beam^{-1}$. $-5\,\sigma$ is shown in white dashed contour.
The ellipses in the bottom left corner of each image indicate the synthesized beams. The black crosses indicate the optical peak.
\label{fig:channalmap}}
\end{figure*}

\begin{figure*}[ht!]
\epsscale{1.2}
\plotone{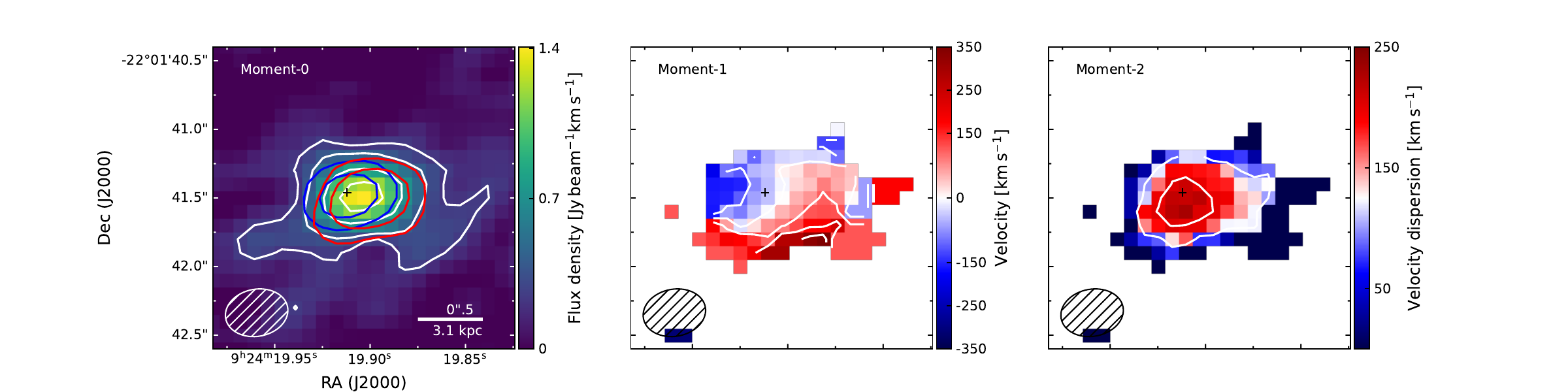}
\caption{\textbf{Left}: Velocity-integrated (moment-0) map of [C\,{\sc ii}]. White contours indicate $3\,\sigma$, $5\,\sigma$, $10\,\sigma$ and $15\,\sigma$, where $1\,\sigma=0.079\,\rm Jy\,beam^{-1}\,km\,s^{-1}$. $-3\,\sigma$ is shown in white dashed contour. Blue and red contours indicate $3\,\sigma$ and $5\,\sigma$ of the blueward and redward parts, respectively.  
\textbf{Middle}: Intensity-weighted velocity (moment-1) map of [C\,{\sc ii}] with $4\,\sigma$ cutoff. White contours indicate $-$300, $-$200, $-$100, 0, 100, 200 and 300 km$\,$s$^{-1}$.
\textbf{Right}: Intensity-weighted velocity dispersion (moment-2) map of [C\,{\sc ii}] with $4\,\sigma$ cutoff. White contours indicate 100 and 200 km$\,$s$^{-1}$.
The ellipses in the bottom left corner of each image indicate the synthesized beams. The black crosses indicate the optical peak.
\label{fig:momentmap}}
\end{figure*}

\begin{figure*}[ht!]
\epsscale{1}
\plotone{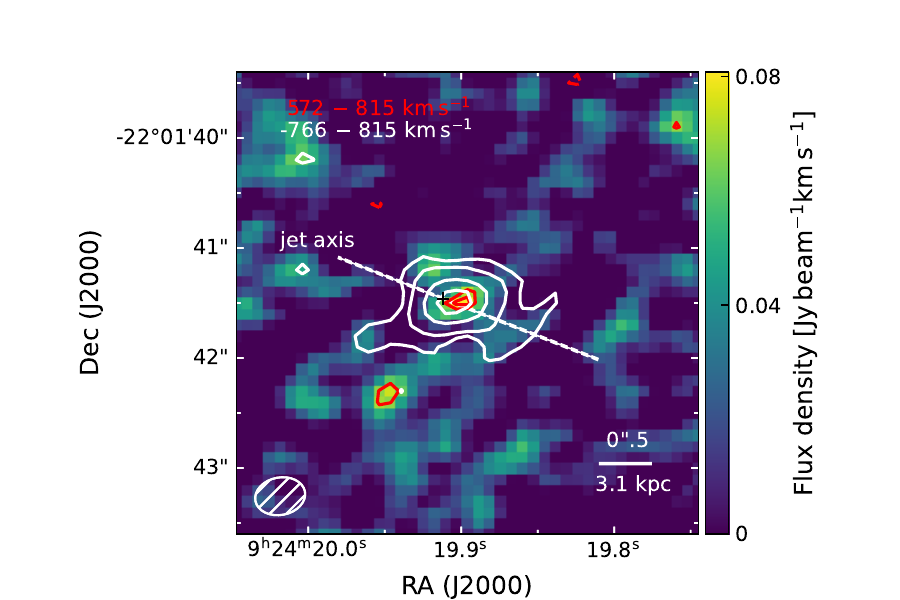}
\caption{The velocity-integrated map of redshifted [C\,{\sc ii}] structure. The red and white texts in the top left corner indicate the integrated velocity range of the redshifted structure and the main component, respectively. 
The red contours indicate $3\,\sigma$, and $3.5\,\sigma$, where $1\,\sigma=0.022\,\rm Jy\,beam^{-1}\,km\,s^{-1}$. $-3\,\sigma$ is shown in red dashed contour. 
White contours indicate $3\,\sigma$, $5\,\sigma$, $10\,\sigma$ and $15\,\sigma$, where $1\,\sigma=0.079\,\rm Jy\,beam^{-1}\,km\,s^{-1}$. $-3\,\sigma$ is shown in white dashed contour.
The ellipse in the bottom left corner indicates the synthesized beam.
The black cross indicates the optical peak. The dashed line indicate the jet axis.
\label{fig:outflow}}
\end{figure*}

From the spectral data cube, we obtained spectra at the positions of TN J0924$-$2201 and three CO(1--0) detected regions. Figure~\ref{fig:4spectra} shows the spectra at these four positions.
The [C\,{\sc ii}] line emission is clearly detected at the position of TN J0924$-$2201. However, there are no detection at three CO(1--0) detected regions. 

For TN J0924$-$2201, the peak of [C\,{\sc ii}] line is apparently offsetting from the Ly$\alpha$-derived redshift $z_{\rm Ly\alpha}$. Applying Gaussian fit on [C\,{\sc ii}] emission, we derived the redshift $z_{\rm [C\,{\sc II}]}=5.1736\pm0.0002$, which is the systemic redshift of TN J0924$-$2201, as [C\,{\sc ii}] is tracing the global ISM in the host galaxy.

Figure~\ref{fig:linefit} shows the zoomed-in views of [C\,{\sc ii}] line, with the relative velocity centered at the peak of [C\,{\sc ii}]. 
The upper left panel shows the spectrum extracted with aperture $\sim0''.8$, which covers the whole galaxy, while the middle left panel shows the spectrum at the peak pixel, whose size is $0''.1\times0''.1$. 
Comparing with the spectrum of the whole galaxy, a redshifted substructure in the spectrum at the peak pixel appears to be more significant.
We also checked the spectrum in the spectral cube which is produced without applying CASA task
{\tt uvcontsub} (Section~\ref{sec:obs}).
In Figure~\ref{fig:linefit}, the bottom left panel shows the spectrum at the peak pixel in the spectral cube without the subtraction of the continuum and the contamination from the nearby bright source. It is clear that the redshifted substructure does still appear.
We applied double-component 1-D Gaussian fit on the spectrum and found a redshifted substructure, which is shown in red dashed line, with a velocity offset of $702\pm17\,\rm km\,s^{-1}$, relative to the main component. 
The fitting results are summarized in Table~\ref{tab:linefit}.

\begin{deluxetable}{lcc}[ht!]
\tablenum{1}
\tablecaption{Gaussian fit of the [C\,\sc ii] \rm emission with $\sim0''8$ aperture \label{tab:linefit}}
\tablecolumns{3}
\tablewidth{0pt}
\tablehead{
\colhead{} & \colhead{Component 1} & \colhead{Component 2}
}
\startdata
Amplitude (mJy$\,$beam$^{-1}$) & $1.01\pm0.02$ & $0.15\pm0.05$\\
Mean (km$\,$s$^{-1}$) & $\pm6$$\tablenotemark{a}$ & $702\pm17$ \\
Velocity dispersion (km$\,$s$^{-1}$) & $270\pm6$ & $47\pm16$ \\
FWHM (km$\,$s$^{-1}$) & $637\pm15$ & $110\pm37$ \\ 
\enddata
\tablecomments{$\tablenotemark{a}$ We define the mean velocity of Component 1 (the main component) as the systemic velocity.}
\end{deluxetable} 

Figure~\ref{fig:linefit} also shows the spectra of C\,{\sc iv}, Ly$\alpha$ and CO(1--0) in the right panels to indicate the velocity offsets to the [C\,{\sc ii}] line.
With respect to the [C\,{\sc ii}]-derived systemic velocity, Ly$\alpha$ \citep[$z_{\rm Ly\alpha}=5.195$;][]{Matsuoka11} is redshifted by $1035\pm10$ $\rm km\,s^{-1}$, which is the largest velocity offset between [C\,{\sc ii}] and Ly$\alpha$ recorded at $z>5$ to date. If we adopt $z_{\rm Ly\alpha}=5.1989$ reported by \cite{Venemans04}, the velocity offset would be even larger ($1223\pm10\,\rm km\,s^{-1}$).
In Figure~\ref{fig:linefit}, the velocities of six Ly$\alpha$ emitters in the overdense region around TN J0924$-$2201 reported by \cite{Venemans04} are also indicated in the panel of Ly$\alpha$.
The systemic redshift of TN J0924$-$2201 is close to the central velocity of the six Ly$\alpha$ emitters which have velocity offset from $-259$ to $474\,\rm km\,s^{-1}$ \citep[$z=5.168$--$5.183$;][]{Venemans04}. 
It should be noted that the narrow-band filter of the Ly$\alpha$ emitter survey covers the velocity range from $-1100$ to $2600\,\rm km\,s^{-1}$ in relative to $z_{\rm [C\,{\sc II}]}$. 
However, there are no Ly$\alpha$ emitters distributed around the $z_{\rm Ly\alpha}$ of TN J0924$-$2201.
The fact that TN J0924$-$2201 is at the central velocity of the Ly$\alpha$ emitters is consistent with other radio galaxies at lower-$z$ \citep{Pentericci00,Kurk04,Venemans02,Best07}.
It may suggest that TN J0924$-$2201 is located in the central region of the overdensity of multiple smaller galaxies, and TN J0924$-$2201 may potentially become the brightest cluster galaxy of a cluster at $z=0$.
Regarding the C\,{\sc iv} emission line, we consider that the possible two peaks appeared in the spectrum correspond to the peaks of C\,{\sc iv} doublet lines, thus the C\,{\sc iv} emission line has intrinsically narrower velocity width, which is consistent with the FWHM of the redshifted [C\,{\sc ii}] component of $\sim100\,\rm km\,s^{-1}$ (Table~\ref{tab:linefit}).
With respect to the systemic redshift, the center of the C\,{\sc iv} doublet lines, rest-frame 1549$\rm \,\AA$ \citep[$z_{\rm C\,{\sc IV}}=5.184$;][]{Matsuoka11} is redshifted by $500\pm10\,\rm km\,s^{-1}$, which is close to the velocity of the redshifted [C\,{\sc ii}] structure.
Besides, CO(1--0) line from the three companions are redshifted by $\sim1500\,\rm km\,s^{-1}$. 

Figure~\ref{fig:channalmap} shows the channel map of the [C\,{\sc ii}] line. The velocity is relative to the systemic velocity. Here we only show images from $-$444$\,\rm km\,s^{-1}$ to 407$\,\rm km\,s^{-1}$, as the signal-to-noise (S/N) ratio of emission per channel is high enough to be recognized ($\rm S/N>5$). Relative to the optical peak, apparently, the [C\,{\sc ii}] emission redward of its peak is located at east. This feature is more obvious after we integrating them into the moment maps.

Figure~\ref{fig:momentmap} shows the velocity-integrated (moment-0) map, intensity-weighted velocity (moment-1) map and intensity-weighted velocity dispersion (moment-2) map of the [C\,{\sc ii}] line emission. 
On the moment-0 map, we also show the blueward and redeward parts, by integrating them separately.
We found that there is a marginal spatial offset $\sim0''.2$ between these two parts, implying rotational motion of the ISM.
Since the moment-0 map is partially resolved, we measured the deconvolved lengths of the major and minor axises of TN J0924$-$2201.
Assuming a symmetric circular disk, we derived its inclination angle of $52.8\pm7.6$ degree.
Applying the 2-D Gaussian fitting on the moment-0 map, the integrated flux of [C\,{\sc ii}] is $3.34\pm0.31\,\rm Jy\,km\,s^{-1}$, and the peak flux is $1.16\pm0.08\,\rm Jy\,beam^{-1}\,km\,s^{-1}$.
We then derived the [C\,{\sc ii}] luminosity $L_{\rm [C\,{\sc II}]}=(2.52\pm0.23)\times10^{9}\,L_{\odot}$.
On the moment-1 map, the velocity gradient is shown in white contours, indicating rotational motion of the ISM.

Furthermore, we also integrated the reddest part, only from 572$\,\rm km\,s^{-1}$ to 815$\,\rm km\,s^{-1}$, to investigate the redshifted structure.
Figure~\ref{fig:outflow} shows that the redshifted structure is concentrated in the central region. This implies that the redshifted structure may related to the activity in the central region.

From the velocity profile and the spatial extent of the [C\,{\sc ii}] emission line, we can derive the dynamical mass with the assumption that [C\,{\sc ii}] emission is tracing the ISM in a virialized system.
Following \cite{Bothwell13}, the virial dynamical mass is:
\begin{equation}
    M_{\rm dyn}^{\rm vir}=1.56\times10^{6}\ \sigma^{2}\ R,
\end{equation}
where $\sigma$ is 1-D velocity dispersion and $R$ is effective radius.
Here we only took the main component of [C\,{\sc ii}] into account.
Given $\sigma=270\pm6\,\rm km\,s^{-1}$ estimated from Gaussian fitting, and $R=1.7\pm0.1\,$kpc, which is the geometric mean of the deconvolved major and minor radii, measured from moment-0 map, we derived $M_{\rm dyn}^{\rm vir}=(1.9\pm0.2)\times10^{11}\,M_{\odot}$, which is consistent with the SED-derived stellar mass $M_{*}=1.3\times10^{11}\,M_{\odot}$ \citep{DeBreuck10}.
The consistency of dynamical mass and stellar mass indicates a gas-poor nature of the system. This is also consistent with the $3\sigma$ upper limit of molecular gas mass $M_{\rm H_{2}}<1.3\times10^{10}\,M_{\odot}$ at the host galaxy \citep{Lee23}.

\section{Discussion} \label{sec:discussion}
\subsection{Large velocity offset of Ly$\alpha$ line} \label{sec:vo}

\begin{figure*}[ht!]
\epsscale{1.2}
\plotone{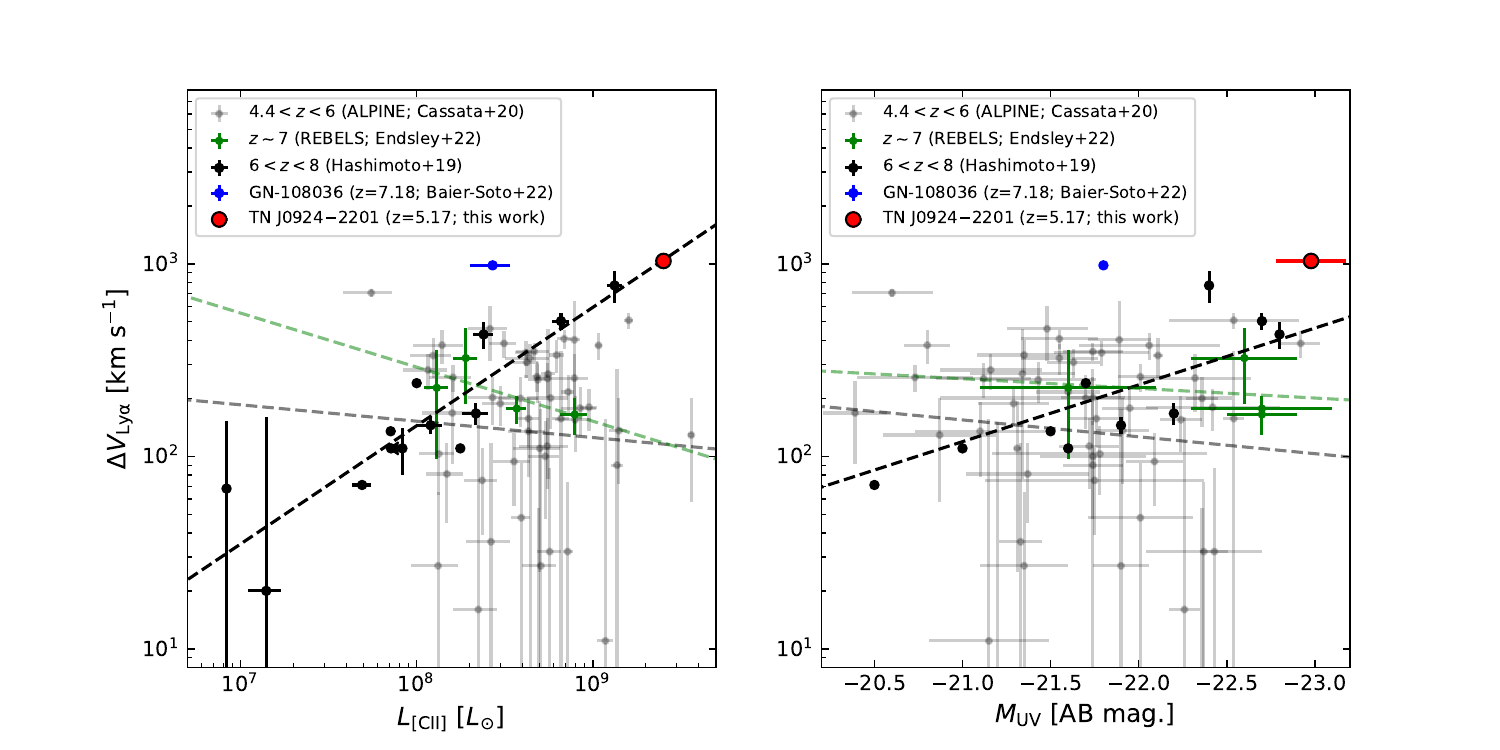}
\caption{\textbf{Left}: Velocity offset between Ly$\alpha$ and [C\,{\sc ii}] versus [C\,{\sc ii}] luminosity. 
\textbf{Right}: Velocity offsets between Ly$\alpha$ and [C\,{\sc ii}] versus UV magnitude. 
The red circle indicates TN J0924$-$2201.
The blue dot indicates the galaxy GN-108036 at $z=7.18$ \citep{Baier-Soto22}.
Gray dots indicate galaxies at $4.4<z<6$ \cite[ALPINE;][]{Cassata20}. Green dots indicate galaxies at $z\sim7$ \cite[REBELS;][]{Endsley22}. Black dots indicate galaxies at $6<z<8$ \citep{Hashimoto19}. The gray, green, and black dashed lines indicate the results of fitting with the ALPINE galaxies at $4.4<z<8$ in \cite{Cassata20}, the REBELS galaxies at $z\sim7$ in \cite{Endsley22}, and the galaxies at $6<z<8$ in \cite{Hashimoto19}, respectively.
\label{fig:dv_LCII_MUV}}
\end{figure*}

\begin{figure}[ht!]
\epsscale{1}
\plotone{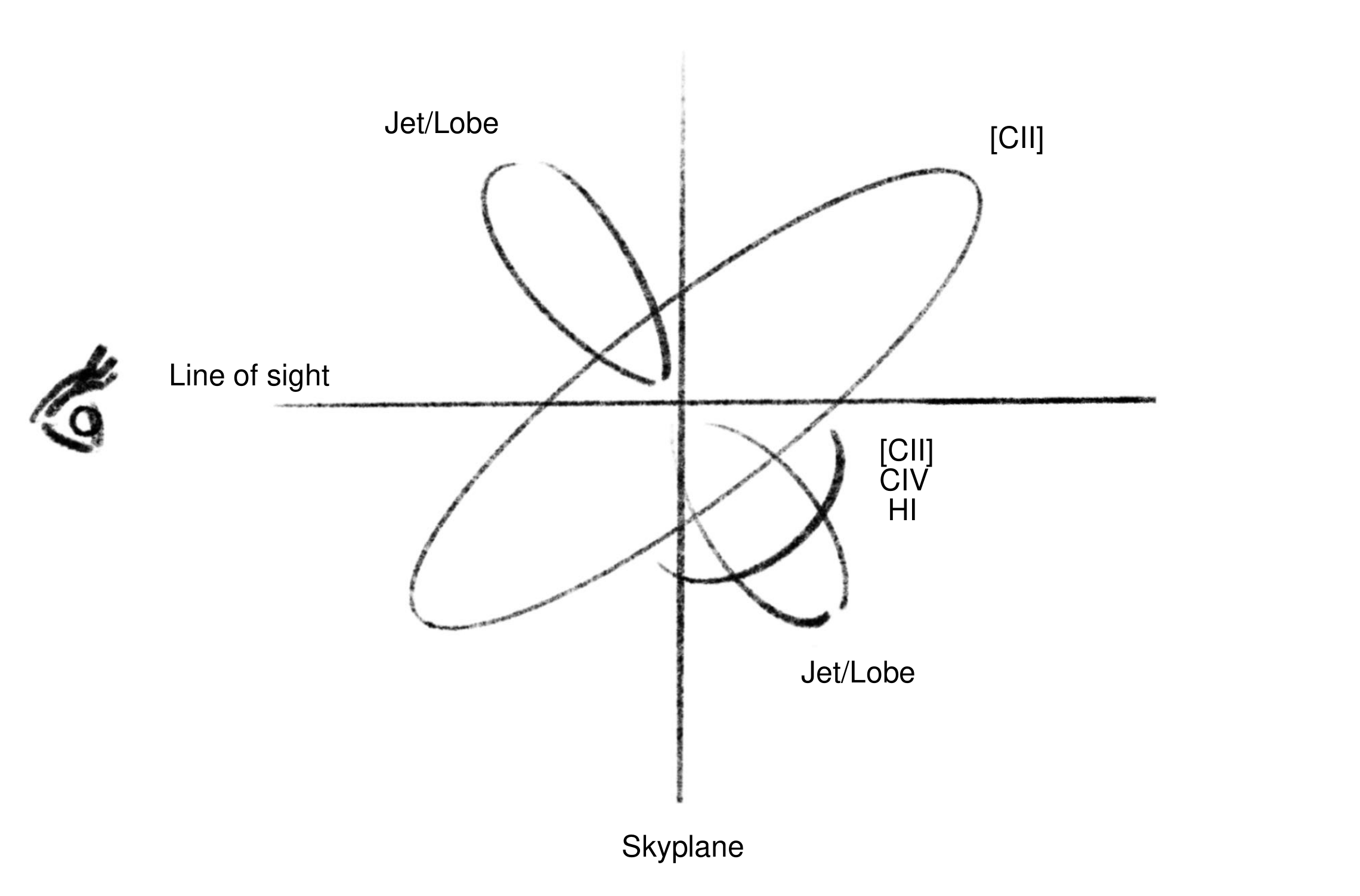}
\caption{Schematic view of TN J0924$-$2201 system. The rotating disk with inclination angle of $52.8\pm7.6$ degree is the main component of [C\,{\sc ii}] line. The observed redshifted [C\,{\sc ii}] and C\,{\sc iv} lines originate from the outflow observed only in one direction.
Neutral hydrogen H\,{\sc i} gas moves with the outflowing [C\,{\sc ii}] and C\,{\sc iv} emitting gas, leading to the large velocity offset of Ly$\alpha$ line.
\label{fig:cartoon}}
\end{figure}

For TN J0924$-$2201, the velocity offset of Ly$\alpha$, $\Delta v_{\rm Ly\alpha}=1035\pm10\,\rm km\,s^{-1}$, is the largest among the velocity offset between [C\,{\sc ii}] and Ly$\alpha$ recorded at $z>5$ to date.
Regarding the velocity offsets between fine structure lines and Ly$\alpha$, 
there are reports of larger velocity offset between [O\,{\sc iii}] and Ly$\alpha$ \citep[$1938\pm162\,\rm km\,s^{-1}$ at $z=8.610$;][]{Tang23}. 
However, based on the large survey of [O\,{\sc iii}] and [C\,{\sc ii}] lines reported by \cite{Cassata20},
[O\,{\sc iii}] tends to be blueshifted relative to [C\,{\sc ii}]. Therefore, here we only discuss the velocity offset between Ly$\alpha$ and [C\,{\sc ii}] lines.

Observationally, Ly$\alpha$ is generally redshifted with respect to the systemic velocity of galaxies.
At high-$z$, the intergalactic medium can absorb most of the blue part of Ly$\alpha$, leading to the redshifted Ly$\alpha$.
While in general, since Ly$\alpha$ is a resonant line, the velocity offset can be modeled with a dynamical structure of Ly$\alpha$ scattering material, and the observed velocity offset can be interpreted with the expanding shell model \citep{Verhamme06}.
In this model, the velocity of the expanding shell is the half of the $\Delta v_{\rm Ly\alpha}$. An alternative interpretation is that galaxies are surrounded by environment with high neutral hydrogen column density $N_{\rm HI}$. If $N_{\rm HI}$ is high, the resonant scattering of Ly$\alpha$ happens multiple times and only highly shifted components of Ly$\alpha$ can escape from the system, resulting in a high $\Delta v_{\rm Ly\alpha}$ and narrow rest-frame equivalent width of Ly$\alpha$ \citep[e.g.,][]{Verhamme15}.

Figure~\ref{fig:dv_LCII_MUV} shows the velocity offset between [C\,{\sc ii}] and Ly$\alpha$, $\Delta v_{\rm Ly\alpha}$, versus [C\,{\sc ii}] luminosity $L_{\rm [C\,{\sc II}]}$, and $\Delta v_{\rm Ly\alpha}$ versus UV magnitude $M_{\rm UV}$. 
We compared TN J0924$-$2201 with case studies from the literature \citep{Hashimoto19,Baier-Soto22} and large ALMA [C\,{\sc ii}] surveys, the ALMA Large Program to INvestigate [C\,{\sc ii}] at Early Times \citep[ALPINE;][]{Cassata20} and the Reionization-Era Bright Emission Line Survey \citep[REBELS;][]{Endsley22}.
We applied linear regression fitting on sample in \cite{Hashimoto19}, ALPINE and REBELS, separately.
The regression analysis of the sample in \cite{Hashimoto19} shows that $\Delta v_{\rm Ly\alpha}$ tightly correlates with $L_{\rm [C\,{\sc II}]}$ and $M_{\rm UV}$.
Whereas, the distribution of ALPINE and REBELS samples does not show such correlation. 
Regarding this discrepancy, firstly, the redshift ranges of the samples are different. ALPINE sample is at $4.4<z<6$ and sample in \cite{Hashimoto19} is at $6<z<8$, which corresponds to the reionization epoch. This may suggest that the discrepancy is caused by the difference in the surrounding environment. 
However, the REBELS sample at $z\sim7$, which also corresponds to the reionization epoch, does not show positive correlation either.
Secondly, noticing that the sample in \cite{Hashimoto19} is collected from the literature instead of selected from a survey, the upcoming complete sample of REBELS may be more appropriate to examine whether the correlation between $\Delta v_{\rm Ly\alpha}$, $L_{\rm [C\,{\sc II}]}$ and $M_{\rm UV}$ exist in the reionization epoch or not.

For TN J0924$-$2201, which is located in an overdense region of Ly$\alpha$ emitters \citep{Venemans04} and Lyman-break galaxies \citep{Overzier06}, the overdense environment may suggest a high hydrogen column density $N_{\rm H}$. However, it is not necessarily related to a high neutral hydrogen column density $N_{\rm HI}$, 
since the process of reionization and recombination is considered to be boosted in an overdense region \citep[e.g.,][]{Choudhury06}. 
Regarding the rest-frame equivalent width of Ly$\alpha$, for galaxies whose $\Delta v_{\rm Ly\alpha}>500\,\rm km\,s^{-1}$, their equivalent widths of Ly$\alpha$ are generally narrower than $50\,\rm \AA$ \citep{Erb14,Nakajima18,Hashimoto19,Cassata20,Endsley22}.
On the contrary, the equivalent width of Ly$\alpha$ \citep[$83^{+148}_{-14}\,\rm \AA$;][]{Venemans04} of TN J0924$-$2201 is not as narrow as those which also have large $\Delta v_{\rm Ly\alpha}$. These observational results appear to contradict with the high $N_{\rm HI}$ scenario for TN J0924$-$2201. 
On the other hand, as mentioned in Section~\ref{sec:cii}, the velocity-integrated image shows that the position of the redshifted [C\,{\sc ii}] structure is close to the center of the galaxy (see Figure~\ref{fig:outflow}), indicating an outflow induced by an activity in the central region, probably the AGN.
If we assume the velocity offset of redshifted [C\,{\sc ii}] structure ($702\pm17\,\rm km\,s^{-1}$) and C\,{\sc iv} ($500\pm10\,\rm km\,s^{-1}$) represent outflow velocity, it is about half of the $\Delta v_{\rm Ly\alpha}$ ($1035\pm10\,\rm km\,s^{-1}$). 
These observational evidences are consistent with the expanding shell model \citep{Verhamme06}, in which a large amount of neutral hydrogen moves with the [C\,{\sc ii}] and C\,{\sc iv} emitting gas (Figure~\ref{fig:cartoon}).
\cite{Verhamme06} also indicate that the Ly$\alpha$ profile from an expanding shell can be broadened by about four times the maximum velocity of the expanding shell. And indeed, the Ly$\alpha$ profile of TN J0924--2201 is broad, extending over about 2000\,km\,s$^{-1}$ (see Figure~\ref{fig:linefit}).

In Figure~\ref{fig:outflow}, although the marginal spatial offset of $\sim0''.2$ between the optical peak and the peak of redshifted [C\,{\sc ii}] structure is smaller than the synthesized beam, we estimated the projected distance of the redshifted [C\,{\sc ii}] from the center of the galaxy to $\sim1\,$kpc. Assuming that the ouflow moves $\sim1\,$kpc with the velocity of $\sim700\,\rm km\,s^{-1}$, the time scale is $\sim1\,$M years, which is comparable with the lifetime of the most massive O-type stars \citep[e.g.,][]{Meynet03}. Therefore, we consider this outflow as an ongoing outflow.

\subsection{Outflowing molecular clouds}

\begin{figure*}[ht!]
\epsscale{1.1}
\plotone{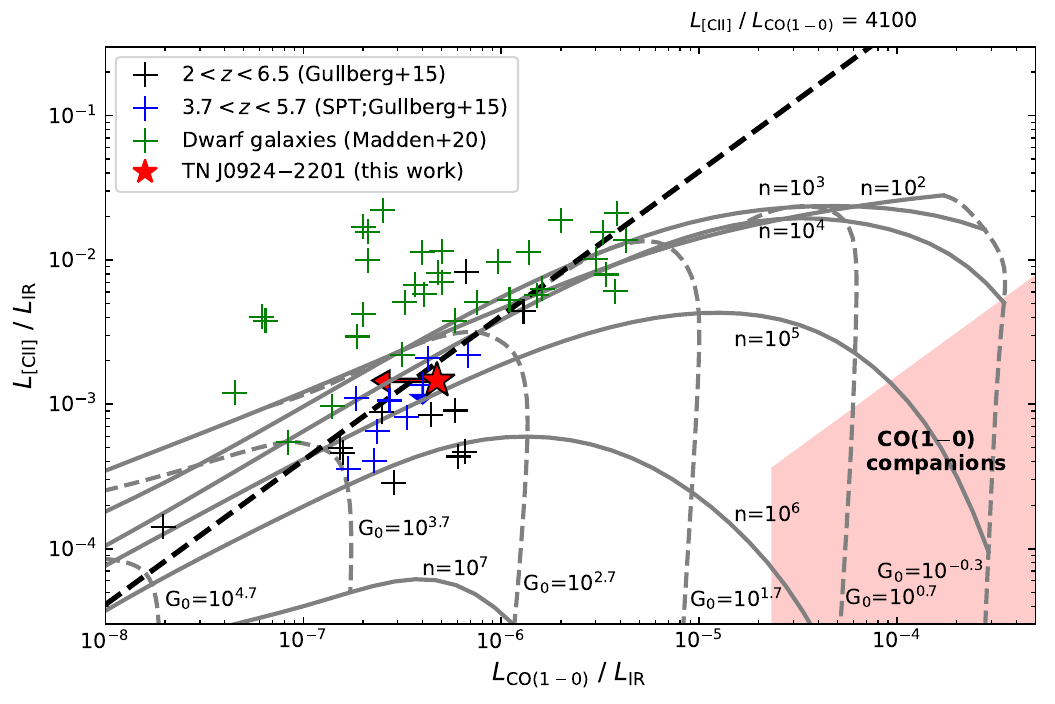}
\caption{$L_{\rm [C\,{\sc II}]}$/$L_{\rm IR}$ versus $L_{\rm CO(1-0)}$/$L_{\rm IR}$. Black crosses indicate galaxies at $2<z<6.5$ \cite[][and references therein]{Gullberg15}. Blue crosses indicate SPT galaxies at $3.7<z<5.7$ \citep{Gullberg15}. Green crosses indicate dwarf galaxies \citep{Madden20}. The red star symbol indicates TN J0924$-$2201. The red shaded area indicates the constrained parameter space of CO(1--0) companions. Gray solid lines and dashed lines indicate parameters in PDR model, $n$ and $G_{0}$, respectively, which are derived by using PDRT \citep{Kaufman06,Pound08,Pound11,Pound23}. The black dashed line indicates line ratio $L_{\rm [C\,{\sc II}]}$/$L_{\rm CO(1-0)}=4100$.
\label{fig:pdr}}
\end{figure*}

\begin{figure}[ht!]
\epsscale{1.2}
\plotone{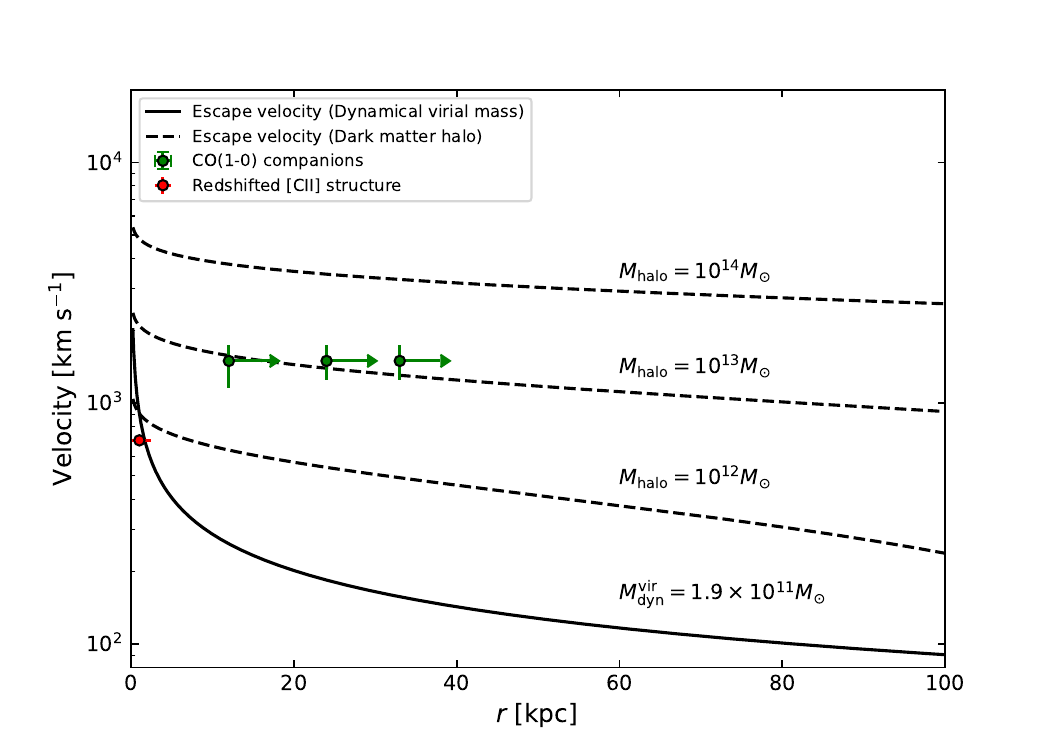}
\caption{Outflow velocity versus distance $r$ from TN J0924$-$2201. 
The solid line indicates escape velocity of a point mass with viral dynamical mass, $M_{\rm dyn}^{\rm vir}=(1.9\pm0.2)\times10^{11}\,M_{\odot}$.
Dashed lines indicate escape velocity of dark matter halos with $M_{\rm halo}=10^{12}\,M_{\odot}$, $M_{\rm halo}=10^{13}\,M_{\odot}$, and $M_{\rm halo}=10^{14}\,M_{\odot}$. The red dot indicates the redshifted [C\,{\sc ii}] structure. The uncertainty of its distance is derived from 0$''$.4, which is the beam size. The green dots indicate CO(1--0) companions. The lower limits of their distances are the projected distances reported by \cite{Lee23}.
\label{fig:escape}}
\end{figure}

Regarding CO(1--0) emission, which is not from the host galaxy of TN J0924$-$2201, the velocity offset between CO(1--0) and the host galaxy is large ($\sim1500\,\rm km\,s^{-1}$). One possibility is that the CO(1--0) companions located at the same redshift with TN J0924$-$2201 but has large velocity offset. In this case, the CO(1--0) companions can possibly be outflow or inflow. Actually, a stream of inflowing [C\,{\sc i}] gas associated with the radio galaxy 4C 41.17 at $z=3.792$ has been reported \citep{Emonts23a}.
As the circumgalactic medium, CO(1--0) companions may brighten the synchrotron luminosity of the close radio lobe because the gas is confined and compressed by the propagating lobe \citep{Emonts23b}. In contrast, the brighter radio lobe of TN J0924$-$2201 is shown at the far side (at east; see Figure~\ref{fig:cont}, VLA 19 GHz image) from the CO(1--0) companions instead of the close side (at west). This can be explained by the Doppler beaming effect as it can be significant enough to surpass the gas-induced brightening. The fainter radio lobe at west may be the counter (receding) jet of TN J0924$-$2201 because the Doppler dimming cancelled the gas-induced brightening. This is consistent with the observed direction of the redshifted outflowing [C\,{\sc ii}] and C\,{\sc iv} emitting gas (Figure~\ref{fig:cartoon}).
However, noticing that the $3\sigma$ upper limit of molecular gas mass constrained by the VLA observation is $1.3\times10^{10}\,M_{\odot}$ with ULIRGs-like conversion factor $\alpha_{\rm CO}$ \citep{Lee23}, it could be about five times higher if the Milky-Way-like conversion factor is adopted.
Except for the host galaxy, it is more reasonable to use the Milky-Way-like conversion factor for estimating the upper limit of molecular gas mass, since only the host galaxy shows dust emission.
Therefore, it is possible that a significant amount of molecular gas also exists on the east side.
Although the molecular gas is likely more abundant on the west side, deeper CO (or other tracers) observations are needed to determine the full distribution of molecular and atomic gas surrounding the host galaxy.

The other possibility is that the CO(1--0) companions are simply located at different redshift, without physical associations with TN J0924$-$2201. In this case, the CO(1--0) companions may originate from dwarf galaxies.

From the current VLA CO(1--0) data, two scenarios appear to be equally plausible. However, the additional ALMA [C\,{\sc ii}] and continuum data allow us to try to distinguish between these two scenarios by constraining the physical conditions of line emitting objects based on the PDR model.
Figure~\ref{fig:pdr} shows the line ratios $L_{\rm [C\,{\sc II}]}$/$L_{\rm IR}$ versus $L_{\rm CO(1-0)}$/$L_{\rm IR}$ of TN J0924$-$2201, CO(1--0) companions and other galaxies \cite[][and references therein]{Madden20, Gullberg15}. 
Gray solid and dashed lines are overplotted to indicate parameters in the PDR model, density $n$ and radiation field strength $G_{0}$, respectively, by using PDRT \citep{Kaufman06,Pound08,Pound11,Pound23}.
Dusty star-forming galaxies selected from the South Pole Telescope (SPT) survey are referred as SPT galaxies \citep{Gullberg15}.
Most of the SPT galaxies distributes close to the diagonal line $L_{\rm [C\,{\sc II}]}$/$L_{\rm CO(1-0)}=4100$. 
The upper limit on $L_{\rm CO(1-0)}$/$L_{\rm IR}$ of TN J0924$-$2201 is still consistent with the distribution of the SPT galaxies. This is consistent with the star-forming nature of TN J0924$-$2201.

For CO(1--0) companions, we derived the upper limit of $L_{\rm [C\,{\sc II}]}$ with $3\sigma$ upper limit of [C\,{\sc ii}] and assuming its velocity width as $500\,\rm km\,s^{-1}$, which is similar with the FWHM of [C\,{\sc ii}] in TN J0924$-$2201.
Given the $3\sigma$ upper limit of 1-mm continuum, we derived the upper limit of $L_{\rm IR}$ by scaling with the gray-body fitting of TN J0924$-$2201.
Utilizing these upper limits, we constrained their range of line ratios (red shaded area in Figure~\ref{fig:pdr}). The diagonal boundary is the upper limit of $L_{\rm [C\,{\sc II}]}$/$L_{\rm CO(1-0)}\sim16$.
As these CO(1--0) companions have abundant molecular gas and low SFRs, one may suggest that their properties are somewhat similar to that of dwarf galaxies.
However, in Figure~\ref{fig:pdr}, neither star-forming galaxies \citep{Gullberg15} nor dwarf galaxies \citep{Madden20} distribute very differently from the CO(1--0) companions.
\cite{Madden20} suggest that dwarf galaxies occupy the ``CO-dark'' region probably because of their low metallicity.
In contrast with that, CO(1--0) companions are located in the ``[C\,{\sc ii}]-dark'' region, indicating their different nature.
Given the detection of CO(1--0) emission line, the metallicity of CO(1--0) companions appears to be higher than that of dwarf galaxies which have not yet experienced significant metal enrichment.
This suggests that the CO(1--0) companions may be outflows because the CO(1--0) companions are more likely to be associated with TN J0924$-$2201, which has already experienced the metallcity evolution during its process of stellar mass assembly.
By applying PDR model, these CO(1--0) companions are likely to have high density ($n>10^{4}\,\rm cm^{-3}$) and in weak radiation field ($G_{0}\lesssim10^{1}$).
High velocity dispersion is one of the possible reasons for such dense CO(1--0) companions to have low SFRs. 

Assuming three CO(1--0)-detected regions are outflows from TN J0924$-$2201, Figure~\ref{fig:escape} shows the comparison of outflow velocities with local escape velocities at different distance from TN J0924$-$2201.
If we only consider the virial dynamical mass as a point mass, CO(1--0) companions can easily exceed the escape velocity.
However, since TN J0924$-$2201 resides at the center of the overdense region, the contribution from the host dark matter halo needs to be taken into account.
Considering a distribution of dark matter with a truncated isothermal sphere, we derived the escape velocity $v_{\rm esc}$ following \cite{Veilleux20}, 
\begin{equation}
    v_{\rm esc}(r)=v_{\rm vir}\sqrt{2\left[1+\ln(\frac{r_{\rm vir}}{r})\right]},
\end{equation}
where $v_{\rm vir}$ is the virial velocity, $r$ is the distance, and $r_{\rm vir}$ is the virial radius.
Regarding different halo masses, we calculated the corresponding $v_{\rm vir}$ and $r_{\rm vir}$ with the formulae in \citet{Bryan98}.
For a $10^{13}\,M_{\odot}$ halo at $z=5.1736$, $v_{\rm vir}\sim621\,\rm km\,s^{-1}$ and $r_{\rm vir}\sim112\,\rm kpc$.
We found that the outflow velocities of CO(1--0) companions at their distances are close to or higher than their local escape velocities of a $10^{13}\,M_{\odot}$ halo, which is the highest halo mass that a protocluster can possibly be assembled at $z\sim5$ \citep[e.g.,][]{Brinch24}.
As Figure~\ref{fig:escape} is indicating, if the physical distances of the three CO(1--0) companions increase, the excess of velocity increases as well.
The lower limits of the distances between TN J0924$-$2201 and CO(1--0) companions are the observed projected distance, because  the projected distances can only be equal to or smaller than the actual physical distances.
Even at projected distances, for two out of the three CO(1--0) companions, we found that their outflow velocities are larger than their local escape velocity of a $10^{13}\,M_{\odot}$ halo.
Therefore, if CO(1--0) companions are outflows, their outflow velocities of $\sim1500\,\rm km\,s^{-1}$ surpass the escape velocities of the system of TN J0924$-$2201.
In other words, we are probably witnessing the removal of massive molecular gas from the system of TN J0924$-$2201.

Although we only have the information of the projected distances and the radial velocities instead of the actual distances and velocities, we roughly estimated the time scale for CO(1--0) companions moving from TN J0924$-$2201 to their present positions.
Assuming the CO(1--0) emitting outflows move through the distance of 12--33 kpc with the velocity of $\sim1500\,\rm km\,s^{-1}$, we estimated the time scale as $\sim10$--20\,M years, which is an order of magnitude longer than that of the outflowing [C\,{\sc ii}] and C\,{\sc iv}.
Therefore, we consider the CO(1--0) companions as the fossil of previous large-scale outflows.

\subsection{Kinetic power of radio jets}

Considering the high velocity offsets of the massive CO(1--0) companions, one may wonder whether the AGN of TN J0924$-$2201 is powerful enough to induce such energetic outflows. 
To estimate the energy balance, we estimated the kinetic power of the possible outflows and the radio jets.

We derived the kinetic power of outflows ($\dot{E_{\rm out}}$) following \cite{Rose18},
\begin{equation}
    \dot{M}=M_{\rm out}\frac{v_{\rm out}}{r}
\end{equation}
and
\begin{equation}
    \dot{E_{\rm out}}=\frac{\dot{M}}{2}\left(v_{\rm out}^{2}+3\sigma^{2}\right),
\end{equation}
where $M_{\rm out}$ is the outflow mass, $v_{\rm out}$ is the outflow velocity, $r$ is the outflow radius and $\sigma$ is the velocity dispersion, assuming Gaussian line profile.
By using the ULIRG-like CO-to-H$_{2}$ conversion factor $\alpha_{\rm CO}$, we estimated $\dot{E_{\rm out}}\sim3.3\times10^{45}\,\rm erg\,s^{-1}$.
However, using ULIRG-like $\alpha_{\rm CO}$ is unrealistic and probably leads to an underestimation. Given the non-detection of 1-mm continuum emission, CO(1--0) companions are not ULIRGs so the actual amount of molecular gas should be more massive.
The estimation of $\dot{E_{\rm out}}$ would be about five times larger if we apply Milky Way-like $\alpha_{\rm CO}$.
Therefore, the estimation here is likely to be a lower limit.

We also derived the kinetic power (kinetic luminosity $L_{\rm kin}$) of radio jets following \cite{Smolcic17}:
\begin{equation}
    L_{\rm kin}(L_{\rm 1.4GHz})=0.86\cdot\log{L_{\rm 1.4GHz}}+14.08+1.5\log{f_{\rm W}},
\end{equation}
where $L_{\rm 1.4GHz}$ is the rest-frame 1.4 GHz radio luminosity and $f_{\rm W}$ is an uncertainty parameter in the range of 1--20. The uncertainty parameter $f_{\rm W}$ represents the different ways of normalization derived from different physics.
For example, $f_{\rm W}=15$ is close to the kinetic luminosity derived from X-ray cavities, which induced by radio jets and lobes in clusters \citep[][and references therein]{Smolcic17}. 
Applying the extreme $f_{\rm W}=20$, we estimated $L_{\rm kin}\sim3.5\times10^{45}\,\rm erg\,s^{-1}$.
Although it appears to be at the same order of magnitude of $\dot{E_{\rm out}}$, we are aware that this is the consistency between the lower limit of $\dot{E_{\rm out}}$ and the upper limit of $L_{\rm kin}$.
The estimation of $L_{\rm kin}$ is about 100 times lower if we apply $f_{\rm W}=1$.

This result indicates that, although TN J0924$-$2201 has a powerful radio AGN, the kinetic power of radio jets alone is barely strong enough to launch the CO(1--0) outflows.
Therefore, the radiative power has to play an important role in launching the outflows even in a powerful radio galaxy like TN J0924$-$2201.
Assuming a typical bolometric luminosity, $L^{*}_{\rm bol}\sim10^{46.5}\,\rm erg\,s^{-1}$, based on the bolometric quasar luminosity function at $z\sim5$ \citep{Shen20}, it appears to be sufficient for launching the estimated $\dot{E_{\rm out}}$.
This is consistent with the unification model that radio-loud quasars are either dormant or hidden in the center of high-$z$ radio galaxies \citep[e.g.,][]{MileyDeBeruck08}.

\subsection{Neutral hydrogen absorber}

Although we interpreted the large velocity offset of Ly$\alpha$ as a result of outflowing gas associated with [C\,{\sc ii}], C\,{\sc iv} and H\,{\sc i}, an alternative interpretation is that the velocity offset is caused by a dusty H\,{\sc i} absorber. This H\,{\sc i} absorber may contain significant amounts of dust, which would absorb the Ly$\alpha$ photons and re-emit them in the FIR rather than resonantly scatter them.
In Figure~\ref{fig:cont}, it appears that there is a possible spatial offset $\sim0''.1$, which is smaller than the ALMA beam size of $0''.38\times0''.28$, between ALMA dust continuum and $HST$ optical peak.
This suggests substantial dust obscuration, which may affect the Ly$\alpha$ profile. In this case, as C\,{\sc iv} is also a resonant line, both the Ly$\alpha$ and C\,{\sc iv} would be offset from the systemic redshift traced by [C\,{\sc ii}] \citep[e.g.,][]{Kolwa19}, although there still is a velocity offset of $\sim500\,$km\,s$^{-1}$ between Ly$\alpha$ and C\,{\sc iv} in TN J0924--2201.
On the other hand, in deep spectroscopy of HzRGs, the rest-frame UV emission lines such as C\,{\sc iv} are generally tracing the systemic velocity of AGN host galaxy, rather than only tracing the outflow \citep[e.g.,][]{Villar-Martin03, Wang23}. While in TN J0924$-$2201, there is a velocity offset of $500\pm10\,\rm km\,s^{-1}$ between C\,{\sc iv} and [C\,{\sc ii}]. Instead of a dynamic outflow, a dusty H\,{\sc i} absorber can also explain the observed large velocity offset.

\section{Conclusions} \label{sec:conclusions}

We conducted ALMA Band-7 observations of TN J0924$-$2201, targeting on the [C\,{\sc ii}] line and the underlying 1-mm continuum emission.
Our main results are the following.

\begin{itemize}
    \item We obtained 1-mm continuum emission of TN J0924$-$2201. However, there is no detection at three CO(1--0)-detected regions. The integrated flux of TN J0924$-$2201 is $829\pm57\,\mu$Jy. We applied modified blackbody fitting on the dusty star formation component of TN J0924$-$2201, with ALMA Band-6 (1.3$\,$mm) and Band-7 (1$\,$mm) data. We derived $T_{\rm dust}=29.31\pm0.4\,$K, $\beta=1.51\pm0.46$ and $L_{\rm IR}=1.72^{+0.13}_{-0.14}\times10^{12}\,L_{\odot}$, indicating that the host galaxy of TN J0924$-$2201 is an ULIRG.
    \item We detected [C\,{\sc ii}] line of TN J0924$-$2201, and derived the systemic redshift $z_{\rm [C\,{\sc II}]}=5.1736\pm0.0002$, which is consistent with the central velocity of six Ly$\alpha$ emitters in the associated overdense region. The systemic redshift also indicates a blueshift than the $z_{\rm Ly\alpha}$ by a velocity offset of $1035\pm10\,\rm km\,s^{-1}$, marking the largest velocity offset between [C\,{\sc ii}] line and Ly$\alpha$ line recorded at $z>5$ to date.
    The integrated flux of [C\,{\sc ii}] is $3.34\pm0.31\,\rm Jy\,km\,s^{-1}$, and $L_{\rm [C\,{\sc II}]}=(2.52\pm0.23)\times10^{9}\,L_{\odot}$. 
    \item The channel map and moment maps indicate that the main component of [C\,{\sc ii}] is a rotational structure. Assuming a symmetric circular disk, we derived its inclination angle of $52\pm7.6$ degree. Also, we derived the virial dynamical mass $M^{\rm vir}_{\rm dyn}=(1.9\pm0.2)\times10^{11}\,M_{\odot}$, which is consistent with the SED-derived stellar mass $M_{*}=1.3\times10^{11} M_{\odot}$.
    \item Applying double-component 1-D Gaussian fit on the spectrum, we discovered a redshifted structure of [C\,{\sc ii}], which has velocity offset of $702\pm17\,\rm km\,s^{-1}$. This is close to the redshift of the UV C\,{\sc iv} emission line, corresponding to the velocity offset of $500\pm10\,\rm km\,s^{-1}$, which is about half of the velocity offset of Ly$\alpha$. By only integrating the velocity range of the redshifted [C\,{\sc ii}] structure, we found that it is concentrated in the central region. These observational results are consistent with a shell outflow induced by the central AGN activity. We estimated the time scale of this outflow as $\sim1\,$M years.
    \item Applying the PDR model with the measured line ratios, our analyses indicate that the nature of three massive CO(1--0) companions is different from the dwarf galaxies which have low metallicity. This suggests that the CO(1--0) companions are possibly associated with TN J0924$-$2201, which has already experienced the metallicity evolution, supporting that they are outflowing molecular gas.
    We estimated the time scale of this outflow as $\sim$10--20$\,$M years.
    Given their high outflow velocity of $\sim1500\,\rm km\,s^{-1}$, the outflowing CO(1--0) companions may exceed the escape velocity of a $10^{13}\,M_{\odot}$ halo, which is the highest halo mass that a protocluster can possibly be assembled at $z\sim5$.
\end{itemize}

Our results collectively suggest that the radio galaxy TN J0924$-$2201, with the ongoing and fossil large-scale outflows, is in a distinctive phase of removal of molecular gas, from a central massive galaxy in an overdense region in the early universe.
An alternative interpretation of the existence of a dusty H\,{\sc i} absorber is also proposed.
While to better understand the nature of the three CO(1--0) companions, deep observations which can provide their kinematic properties are crucial.

\begin{acknowledgments}

We thank the anonymous reviewer for the constructive comments.
This paper makes use of the following ALMA data: ADS/JAO.ALMA\#2021.1.00219.S. ALMA is a partnership of ESO (representing its member states), NSF (USA) and NINS (Japan), together with NRC (Canada), MOST and ASIAA (Taiwan), and KASI (Republic of Korea), in cooperation with the Republic of Chile. The Joint ALMA Observatory is operated by ESO, AUI/NRAO and NAOJ.
Data analysis was in part carried out on the Multi-wavelength Data Analysis System operated by the Astronomy Data Center (ADC), National Astronomical Observatory of Japan.
This work is supported by NAOJ ALMA Scientific Research Grant Code 2023-24A.
This work is supported by JSPS KAKENHI grant No. JP21H01133, JP17H06130, JP22H04939, and JP23K20035.

\end{acknowledgments}


\vspace{5mm}
\facilities{ALMA}


\software{Astropy \citep{astropy13,astropy18,astropy22},  
          CASA \citep{CASA22},
          PDRT \citep{Kaufman06,Pound08,Pound11,Pound23},
          SciPy \citep{SciPy20}
          }



%\appendix

%\section{Appendix information}

\bibliography{Mybib}{}

\begin{thebibliography}{}
\expandafter\ifx\csname natexlab\endcsname\relax\def\natexlab#1{#1}\fi
\providecommand{\url}[1]{\href{#1}{#1}}
\providecommand{\dodoi}[1]{doi:~\href{http://doi.org/#1}{\nolinkurl{#1}}}
\providecommand{\doeprint}[1]{\href{http://ascl.net/#1}{\nolinkurl{http://ascl.net/#1}}}
\providecommand{\doarXiv}[1]{\href{https://arxiv.org/abs/#1}{\nolinkurl{https://arxiv.org/abs/#1}}}

\bibitem[{{Astropy Collaboration} {et~al.}(2013){Astropy Collaboration}, {Robitaille}, {Tollerud}, {Greenfield}, {Droettboom}, {Bray}, {Aldcroft}, {Davis}, {Ginsburg}, {Price-Whelan}, {Kerzendorf}, {Conley}, {Crighton}, {Barbary}, {Muna}, {Ferguson}, {Grollier}, {Parikh}, {Nair}, {Unther}, {Deil}, {Woillez}, {Conseil}, {Kramer}, {Turner}, {Singer}, {Fox}, {Weaver}, {Zabalza}, {Edwards}, {Azalee Bostroem}, {Burke}, {Casey}, {Crawford}, {Dencheva}, {Ely}, {Jenness}, {Labrie}, {Lim}, {Pierfederici}, {Pontzen}, {Ptak}, {Refsdal}, {Servillat}, \& {Streicher}}]{astropy13}
{Astropy Collaboration}, {Robitaille}, T.~P., {Tollerud}, E.~J., {et~al.} 2013, \aap, 558, A33, \dodoi{10.1051/0004-6361/201322068}

\bibitem[{{Astropy Collaboration} {et~al.}(2018){Astropy Collaboration}, {Price-Whelan}, {Sip{\H{o}}cz}, {G{\"u}nther}, {Lim}, {Crawford}, {Conseil}, {Shupe}, {Craig}, {Dencheva}, {Ginsburg}, {Vand erPlas}, {Bradley}, {P{\'e}rez-Su{\'a}rez}, {de Val-Borro}, {Aldcroft}, {Cruz}, {Robitaille}, {Tollerud}, {Ardelean}, {Babej}, {Bach}, {Bachetti}, {Bakanov}, {Bamford}, {Barentsen}, {Barmby}, {Baumbach}, {Berry}, {Biscani}, {Boquien}, {Bostroem}, {Bouma}, {Brammer}, {Bray}, {Breytenbach}, {Buddelmeijer}, {Burke}, {Calderone}, {Cano Rodr{\'\i}guez}, {Cara}, {Cardoso}, {Cheedella}, {Copin}, {Corrales}, {Crichton}, {D'Avella}, {Deil}, {Depagne}, {Dietrich}, {Donath}, {Droettboom}, {Earl}, {Erben}, {Fabbro}, {Ferreira}, {Finethy}, {Fox}, {Garrison}, {Gibbons}, {Goldstein}, {Gommers}, {Greco}, {Greenfield}, {Groener}, {Grollier}, {Hagen}, {Hirst}, {Homeier}, {Horton}, {Hosseinzadeh}, {Hu}, {Hunkeler}, {Ivezi{\'c}}, {Jain}, {Jenness}, {Kanarek}, {Kendrew}, {Kern}, {Kerzendorf}, {Khvalko}, {King}, {Kirkby}, {Kulkarni},
  {Kumar}, {Lee}, {Lenz}, {Littlefair}, {Ma}, {Macleod}, {Mastropietro}, {McCully}, {Montagnac}, {Morris}, {Mueller}, {Mumford}, {Muna}, {Murphy}, {Nelson}, {Nguyen}, {Ninan}, {N{\"o}the}, {Ogaz}, {Oh}, {Parejko}, {Parley}, {Pascual}, {Patil}, {Patil}, {Plunkett}, {Prochaska}, {Rastogi}, {Reddy Janga}, {Sabater}, {Sakurikar}, {Seifert}, {Sherbert}, {Sherwood-Taylor}, {Shih}, {Sick}, {Silbiger}, {Singanamalla}, {Singer}, {Sladen}, {Sooley}, {Sornarajah}, {Streicher}, {Teuben}, {Thomas}, {Tremblay}, {Turner}, {Terr{\'o}n}, {van Kerkwijk}, {de la Vega}, {Watkins}, {Weaver}, {Whitmore}, {Woillez}, {Zabalza}, \& {Astropy Contributors}}]{astropy18}
{Astropy Collaboration}, {Price-Whelan}, A.~M., {Sip{\H{o}}cz}, B.~M., {et~al.} 2018, \aj, 156, 123, \dodoi{10.3847/1538-3881/aabc4f}

\bibitem[{{Astropy Collaboration} {et~al.}(2022){Astropy Collaboration}, {Price-Whelan}, {Lim}, {Earl}, {Starkman}, {Bradley}, {Shupe}, {Patil}, {Corrales}, {Brasseur}, {N{"o}the}, {Donath}, {Tollerud}, {Morris}, {Ginsburg}, {Vaher}, {Weaver}, {Tocknell}, {Jamieson}, {van Kerkwijk}, {Robitaille}, {Merry}, {Bachetti}, {G{"u}nther}, {Aldcroft}, {Alvarado-Montes}, {Archibald}, {B{'o}di}, {Bapat}, {Barentsen}, {Baz{'a}n}, {Biswas}, {Boquien}, {Burke}, {Cara}, {Cara}, {Conroy}, {Conseil}, {Craig}, {Cross}, {Cruz}, {D'Eugenio}, {Dencheva}, {Devillepoix}, {Dietrich}, {Eigenbrot}, {Erben}, {Ferreira}, {Foreman-Mackey}, {Fox}, {Freij}, {Garg}, {Geda}, {Glattly}, {Gondhalekar}, {Gordon}, {Grant}, {Greenfield}, {Groener}, {Guest}, {Gurovich}, {Handberg}, {Hart}, {Hatfield-Dodds}, {Homeier}, {Hosseinzadeh}, {Jenness}, {Jones}, {Joseph}, {Kalmbach}, {Karamehmetoglu}, {Ka{l}uszy{'n}ski}, {Kelley}, {Kern}, {Kerzendorf}, {Koch}, {Kulumani}, {Lee}, {Ly}, {Ma}, {MacBride}, {Maljaars}, {Muna}, {Murphy}, {Norman}, {O'Steen},
  {Oman}, {Pacifici}, {Pascual}, {Pascual-Granado}, {Patil}, {Perren}, {Pickering}, {Rastogi}, {Roulston}, {Ryan}, {Rykoff}, {Sabater}, {Sakurikar}, {Salgado}, {Sanghi}, {Saunders}, {Savchenko}, {Schwardt}, {Seifert-Eckert}, {Shih}, {Jain}, {Shukla}, {Sick}, {Simpson}, {Singanamalla}, {Singer}, {Singhal}, {Sinha}, {Sip{H{o}}cz}, {Spitler}, {Stansby}, {Streicher}, {{{S}}umak}, {Swinbank}, {Taranu}, {Tewary}, {Tremblay}, {Val-Borro}, {Van Kooten}, {Vasovi{'c}}, {Verma}, {de Miranda Cardoso}, {Williams}, {Wilson}, {Winkel}, {Wood-Vasey}, {Xue}, {Yoachim}, {Zhang}, {Zonca}, \& {Astropy Project Contributors}}]{astropy22}
{Astropy Collaboration}, {Price-Whelan}, A.~M., {Lim}, P.~L., {et~al.} 2022, apj, 935, 167, \dodoi{10.3847/1538-4357/ac7c74}

\bibitem[{{Baier-Soto} {et~al.}(2022){Baier-Soto}, {Herrera-Camus}, {F{\"o}rster Schreiber}, {Contursi}, {Genzel}, {Lutz}, \& {Tacconi}}]{Baier-Soto22}
{Baier-Soto}, R., {Herrera-Camus}, R., {F{\"o}rster Schreiber}, N.~M., {et~al.} 2022, \aap, 664, L5, \dodoi{10.1051/0004-6361/202243642}

\bibitem[{{Best} {et~al.}(2007){Best}, {von der Linden}, {Kauffmann}, {Heckman}, \& {Kaiser}}]{Best07}
{Best}, P.~N., {von der Linden}, A., {Kauffmann}, G., {Heckman}, T.~M., \& {Kaiser}, C.~R. 2007, \mnras, 379, 894, \dodoi{10.1111/j.1365-2966.2007.11937.x}

\bibitem[{{Bothwell} {et~al.}(2013){Bothwell}, {Smail}, {Chapman}, {Genzel}, {Ivison}, {Tacconi}, {Alaghband-Zadeh}, {Bertoldi}, {Blain}, {Casey}, {Cox}, {Greve}, {Lutz}, {Neri}, {Omont}, \& {Swinbank}}]{Bothwell13}
{Bothwell}, M.~S., {Smail}, I., {Chapman}, S.~C., {et~al.} 2013, \mnras, 429, 3047, \dodoi{10.1093/mnras/sts562}

\bibitem[{{Brinch} {et~al.}(2024){Brinch}, {Greve}, {Sanders}, {McPartland}, {Chartab}, {Gillman}, {Vijayan}, {Lee}, {Brammer}, {Casey}, {Ilbert}, {Jin}, {Magdis}, {McCracken}, {Sillassen}, {Toft}, \& {Zavala}}]{Brinch24}
{Brinch}, M., {Greve}, T.~R., {Sanders}, D.~B., {et~al.} 2024, \mnras, 527, 6591, \dodoi{10.1093/mnras/stad3409}

\bibitem[{{Bryan} \& {Norman}(1998)}]{Bryan98}
{Bryan}, G.~L., \& {Norman}, M.~L. 1998, \apj, 495, 80, \dodoi{10.1086/305262}

\bibitem[{{CASA Team} {et~al.}(2022){CASA Team}, {Bean}, {Bhatnagar}, {Castro}, {Donovan Meyer}, {Emonts}, {Garcia}, {Garwood}, {Golap}, {Gonzalez Villalba}, {Harris}, {Hayashi}, {Hoskins}, {Hsieh}, {Jagannathan}, {Kawasaki}, {Keimpema}, {Kettenis}, {Lopez}, {Marvil}, {Masters}, {McNichols}, {Mehringer}, {Miel}, {Moellenbrock}, {Montesino}, {Nakazato}, {Ott}, {Petry}, {Pokorny}, {Raba}, {Rau}, {Schiebel}, {Schweighart}, {Sekhar}, {Shimada}, {Small}, {Steeb}, {Sugimoto}, {Suoranta}, {Tsutsumi}, {van Bemmel}, {Verkouter}, {Wells}, {Xiong}, {Szomoru}, {Griffith}, {Glendenning}, \& {Kern}}]{CASA22}
{CASA Team}, {Bean}, B., {Bhatnagar}, S., {et~al.} 2022, \pasp, 134, 114501, \dodoi{10.1088/1538-3873/ac9642}

\bibitem[{{Casey}(2012)}]{Casey12}
{Casey}, C.~M. 2012, \mnras, 425, 3094, \dodoi{10.1111/j.1365-2966.2012.21455.x}

\bibitem[{{Casey} {et~al.}(2014){Casey}, {Narayanan}, \& {Cooray}}]{Casey14}
{Casey}, C.~M., {Narayanan}, D., \& {Cooray}, A. 2014, \physrep, 541, 45, \dodoi{10.1016/j.physrep.2014.02.009}

\bibitem[{{Cassata} {et~al.}(2020){Cassata}, {Morselli}, {Faisst}, {Ginolfi}, {B{\'e}thermin}, {Capak}, {Le F{\`e}vre}, {Schaerer}, {Silverman}, {Yan}, {Lemaux}, {Romano}, {Talia}, {Bardelli}, {Boquien}, {Cimatti}, {Dessauges-Zavadsky}, {Fudamoto}, {Fujimoto}, {Giavalisco}, {Hathi}, {Ibar}, {Jones}, {Koekemoer}, {M{\'e}ndez-Hernandez}, {Mancini}, {Oesch}, {Pozzi}, {Riechers}, {Rodighiero}, {Vergani}, {Zamorani}, \& {Zucca}}]{Cassata20}
{Cassata}, P., {Morselli}, L., {Faisst}, A., {et~al.} 2020, \aap, 643, A6, \dodoi{10.1051/0004-6361/202037517}

\bibitem[{{Choudhury} {et~al.}(2009){Choudhury}, {Haehnelt}, \& {Regan}}]{Choudhury06}
{Choudhury}, T.~R., {Haehnelt}, M.~G., \& {Regan}, J. 2009, \mnras, 394, 960, \dodoi{10.1111/j.1365-2966.2008.14383.x}

\bibitem[{{Conley} {et~al.}(2011){Conley}, {Cooray}, {Vieira}, {Gonz{\'a}lez Solares}, {Kim}, {Aguirre}, {Amblard}, {Auld}, {Baker}, {Beelen}, {Blain}, {Blundell}, {Bock}, {Bradford}, {Bridge}, {Brisbin}, {Burgarella}, {Carpenter}, {Chanial}, {Chapin}, {Christopher}, {Clements}, {Cox}, {Djorgovski}, {Dowell}, {Eales}, {Earle}, {Ellsworth-Bowers}, {Farrah}, {Franceschini}, {Frayer}, {Fu}, {Gavazzi}, {Glenn}, {Griffin}, {Gurwell}, {Halpern}, {Ibar}, {Ivison}, {Jarvis}, {Kamenetzky}, {Krips}, {Levenson}, {Lupu}, {Mahabal}, {Maloney}, {Maraston}, {Marchetti}, {Marsden}, {Matsuhara}, {Mortier}, {Murphy}, {Naylor}, {Neri}, {Nguyen}, {Oliver}, {Omont}, {Page}, {Papageorgiou}, {Pearson}, {P{\'e}rez-Fournon}, {Pohlen}, {Rangwala}, {Rawlings}, {Raymond}, {Riechers}, {Rodighiero}, {Roseboom}, {Rowan-Robinson}, {Schulz}, {Scott}, {Scott}, {Serra}, {Seymour}, {Shupe}, {Smith}, {Symeonidis}, {Tugwell}, {Vaccari}, {Valiante}, {Valtchanov}, {Verma}, {Viero}, {Vigroux}, {Wang}, {Wiebe}, {Wright}, {Xu}, {Zeimann}, {Zemcov}, \&
  {Zmuidzinas}}]{Conley11}
{Conley}, A., {Cooray}, A., {Vieira}, J.~D., {et~al.} 2011, \apjl, 732, L35, \dodoi{10.1088/2041-8205/732/2/L35}

\bibitem[{{De Breuck} {et~al.}(2000){De Breuck}, {van Breugel}, {R{\"o}ttgering}, \& {Miley}}]{DeBreuck00}
{De Breuck}, C., {van Breugel}, W., {R{\"o}ttgering}, H.~J.~A., \& {Miley}, G. 2000, \aaps, 143, 303, \dodoi{10.1051/aas:2000181}

\bibitem[{{De Breuck} {et~al.}(2010){De Breuck}, {Seymour}, {Stern}, {Willner}, {Eisenhardt}, {Fazio}, {Galametz}, {Lacy}, {Rettura}, {Rocca-Volmerange}, \& {Vernet}}]{DeBreuck10}
{De Breuck}, C., {Seymour}, N., {Stern}, D., {et~al.} 2010, \apj, 725, 36, \dodoi{10.1088/0004-637X/725/1/36}

\bibitem[{{Drouart} {et~al.}(2014){Drouart}, {De Breuck}, {Vernet}, {Seymour}, {Lehnert}, {Barthel}, {Bauer}, {Ibar}, {Galametz}, {Haas}, {Hatch}, {Mullaney}, {Nesvadba}, {Rocca-Volmerange}, {R{\"o}ttgering}, {Stern}, \& {Wylezalek}}]{Drouart14}
{Drouart}, G., {De Breuck}, C., {Vernet}, J., {et~al.} 2014, \aap, 566, A53, \dodoi{10.1051/0004-6361/201323310}

\bibitem[{{Emonts} {et~al.}(2014){Emonts}, {Norris}, {Feain}, {Mao}, {Ekers}, {Miley}, {Seymour}, {R{\"o}ttgering}, {Villar-Mart{\'\i}n}, {Sadler}, {Carilli}, {Mahony}, {de Breuck}, {Stroe}, {Pentericci}, {van Moorsel}, {Drouart}, {Ivison}, {Greve}, {Humphrey}, {Wylezalek}, \& {Tadhunter}}]{Emonts14}
{Emonts}, B.~H.~C., {Norris}, R.~P., {Feain}, I., {et~al.} 2014, \mnras, 438, 2898, \dodoi{10.1093/mnras/stt2398}

\bibitem[{{Emonts} {et~al.}(2023{\natexlab{a}}){Emonts}, {Lehnert}, {Yoon}, {Mandelker}, {Villar-Mart{\'\i}n}, {Miley}, {De Breuck}, {P{\'e}rez-Torres}, {Hatch}, \& {Guillard}}]{Emonts23a}
{Emonts}, B. H.~C., {Lehnert}, M.~D., {Yoon}, I., {et~al.} 2023{\natexlab{a}}, Science, 379, 1323, \dodoi{10.1126/science.abh2150}

\bibitem[{{Emonts} {et~al.}(2023{\natexlab{b}}){Emonts}, {Lehnert}, {Lebowitz}, {Miley}, {Villar-Mart{\'\i}n}, {Norris}, {De Breuck}, {Carilli}, \& {Feain}}]{Emonts23b}
{Emonts}, B. H.~C., {Lehnert}, M.~D., {Lebowitz}, S., {et~al.} 2023{\natexlab{b}}, \apj, 952, 148, \dodoi{10.3847/1538-4357/acde53}

\bibitem[{{Endsley} {et~al.}(2022){Endsley}, {Stark}, {Bouwens}, {Schouws}, {Smit}, {Stefanon}, {Inami}, {Bowler}, {Oesch}, {Gonzalez}, {Aravena}, {da Cunha}, {Dayal}, {Ferrara}, {Graziani}, {Nanayakkara}, {Pallottini}, {Schneider}, {Sommovigo}, {Topping}, {van der Werf}, \& {Hutter}}]{Endsley22}
{Endsley}, R., {Stark}, D.~P., {Bouwens}, R.~J., {et~al.} 2022, \mnras, 517, 5642, \dodoi{10.1093/mnras/stac3064}

\bibitem[{{Erb} {et~al.}(2014){Erb}, {Steidel}, {Trainor}, {Bogosavljevi{\'c}}, {Shapley}, {Nestor}, {Kulas}, {Law}, {Strom}, {Rudie}, {Reddy}, {Pettini}, {Konidaris}, {Mace}, {Matthews}, \& {McLean}}]{Erb14}
{Erb}, D.~K., {Steidel}, C.~C., {Trainor}, R.~F., {et~al.} 2014, \apj, 795, 33, \dodoi{10.1088/0004-637X/795/1/33}

\bibitem[{{Falkendal} {et~al.}(2019){Falkendal}, {De Breuck}, {Lehnert}, {Drouart}, {Vernet}, {Emonts}, {Lee}, {Nesvadba}, {Seymour}, {B{\'e}thermin}, {Kolwa}, {Gullberg}, \& {Wylezalek}}]{Falkendal19}
{Falkendal}, T., {De Breuck}, C., {Lehnert}, M.~D., {et~al.} 2019, \aap, 621, A27, \dodoi{10.1051/0004-6361/201732485}

\bibitem[{{Gaia Collaboration} {et~al.}(2021){Gaia Collaboration}, {Brown}, {Vallenari}, {Prusti}, {de Bruijne}, {Babusiaux}, {Biermann}, {Creevey}, {Evans}, {Eyer}, {Hutton}, {Jansen}, {Jordi}, {Klioner}, {Lammers}, {Lindegren}, {Luri}, {Mignard}, {Panem}, {Pourbaix}, {Randich}, {Sartoretti}, {Soubiran}, {Walton}, {Arenou}, {Bailer-Jones}, {Bastian}, {Cropper}, {Drimmel}, {Katz}, {Lattanzi}, {van Leeuwen}, {Bakker}, {Cacciari}, {Casta{\~n}eda}, {De Angeli}, {Ducourant}, {Fabricius}, {Fouesneau}, {Fr{\'e}mat}, {Guerra}, {Guerrier}, {Guiraud}, {Jean-Antoine Piccolo}, {Masana}, {Messineo}, {Mowlavi}, {Nicolas}, {Nienartowicz}, {Pailler}, {Panuzzo}, {Riclet}, {Roux}, {Seabroke}, {Sordo}, {Tanga}, {Th{\'e}venin}, {Gracia-Abril}, {Portell}, {Teyssier}, {Altmann}, {Andrae}, {Bellas-Velidis}, {Benson}, {Berthier}, {Blomme}, {Brugaletta}, {Burgess}, {Busso}, {Carry}, {Cellino}, {Cheek}, {Clementini}, {Damerdji}, {Davidson}, {Delchambre}, {Dell'Oro}, {Fern{\'a}ndez-Hern{\'a}ndez}, {Galluccio}, {Garc{\'\i}a-Lario},
  {Garcia-Reinaldos}, {Gonz{\'a}lez-N{\'u}{\~n}ez}, {Gosset}, {Haigron}, {Halbwachs}, {Hambly}, {Harrison}, {Hatzidimitriou}, {Heiter}, {Hern{\'a}ndez}, {Hestroffer}, {Hodgkin}, {Holl}, {Jan{\ss}en}, {Jevardat de Fombelle}, {Jordan}, {Krone-Martins}, {Lanzafame}, {L{\"o}ffler}, {Lorca}, {Manteiga}, {Marchal}, {Marrese}, {Moitinho}, {Mora}, {Muinonen}, {Osborne}, {Pancino}, {Pauwels}, {Petit}, {Recio-Blanco}, {Richards}, {Riello}, {Rimoldini}, {Robin}, {Roegiers}, {Rybizki}, {Sarro}, {Siopis}, {Smith}, {Sozzetti}, {Ulla}, {Utrilla}, {van Leeuwen}, {van Reeven}, {Abbas}, {Abreu Aramburu}, {Accart}, {Aerts}, {Aguado}, {Ajaj}, {Altavilla}, {{\'A}lvarez}, {{\'A}lvarez Cid-Fuentes}, {Alves}, {Anderson}, {Anglada Varela}, {Antoja}, {Audard}, {Baines}, {Baker}, {Balaguer-N{\'u}{\~n}ez}, {Balbinot}, {Balog}, {Barache}, {Barbato}, {Barros}, {Barstow}, {Bartolom{\'e}}, {Bassilana}, {Bauchet}, {Baudesson-Stella}, {Becciani}, {Bellazzini}, {Bernet}, {Bertone}, {Bianchi}, {Blanco-Cuaresma}, {Boch}, {Bombrun}, {Bossini},
  {Bouquillon}, {Bragaglia}, {Bramante}, {Breedt}, {Bressan}, {Brouillet}, {Bucciarelli}, {Burlacu}, {Busonero}, {Butkevich}, {Buzzi}, {Caffau}, {Cancelliere}, {C{\'a}novas}, {Cantat-Gaudin}, {Carballo}, {Carlucci}, {Carnerero}, {Carrasco}, {Casamiquela}, {Castellani}, {Castro-Ginard}, {Castro Sampol}, {Chaoul}, {Charlot}, {Chemin}, {Chiavassa}, {Cioni}, {Comoretto}, {Cooper}, {Cornez}, {Cowell}, {Crifo}, {Crosta}, {Crowley}, {Dafonte}, {Dapergolas}, {David}, {David}, {de Laverny}, {De Luise}, {De March}, {De Ridder}, {de Souza}, {de Teodoro}, {de Torres}, {del Peloso}, {del Pozo}, {Delbo}, {Delgado}, {Delgado}, {Delisle}, {Di Matteo}, {Diakite}, {Diener}, {Distefano}, {Dolding}, {Eappachen}, {Edvardsson}, {Enke}, {Esquej}, {Fabre}, {Fabrizio}, {Faigler}, {Fedorets}, {Fernique}, {Fienga}, {Figueras}, {Fouron}, {Fragkoudi}, {Fraile}, {Franke}, {Gai}, {Garabato}, {Garcia-Gutierrez}, {Garc{\'\i}a-Torres}, {Garofalo}, {Gavras}, {Gerlach}, {Geyer}, {Giacobbe}, {Gilmore}, {Girona}, {Giuffrida}, {Gomel}, {Gomez},
  {Gonzalez-Santamaria}, {Gonz{\'a}lez-Vidal}, {Granvik}, {Guti{\'e}rrez-S{\'a}nchez}, {Guy}, {Hauser}, {Haywood}, {Helmi}, {Hidalgo}, {Hilger}, {H{\l}adczuk}, {Hobbs}, {Holland}, {Huckle}, {Jasniewicz}, {Jonker}, {Juaristi Campillo}, {Julbe}, {Karbevska}, {Kervella}, {Khanna}, {Kochoska}, {Kontizas}, {Kordopatis}, {Korn}, {Kostrzewa-Rutkowska}, {Kruszy{\'n}ska}, {Lambert}, {Lanza}, {Lasne}, {Le Campion}, {Le Fustec}, {Lebreton}, {Lebzelter}, {Leccia}, {Leclerc}, {Lecoeur-Taibi}, {Liao}, {Licata}, {Lindstr{\o}m}, {Lister}, {Livanou}, {Lobel}, {Madrero Pardo}, {Managau}, {Mann}, {Marchant}, {Marconi}, {Marcos Santos}, {Marinoni}, {Marocco}, {Marshall}, {Martin Polo}, {Mart{\'\i}n-Fleitas}, {Masip}, {Massari}, {Mastrobuono-Battisti}, {Mazeh}, {McMillan}, {Messina}, {Michalik}, {Millar}, {Mints}, {Molina}, {Molinaro}, {Moln{\'a}r}, {Montegriffo}, {Mor}, {Morbidelli}, {Morel}, {Morris}, {Mulone}, {Munoz}, {Muraveva}, {Murphy}, {Musella}, {Noval}, {Ord{\'e}novic}, {Orr{\`u}}, {Osinde}, {Pagani}, {Pagano},
  {Palaversa}, {Palicio}, {Panahi}, {Pawlak}, {Pe{\~n}alosa Esteller}, {Penttil{\"a}}, {Piersimoni}, {Pineau}, {Plachy}, {Plum}, {Poggio}, {Poretti}, {Poujoulet}, {Pr{\v{s}}a}, {Pulone}, {Racero}, {Ragaini}, {Rainer}, {Raiteri}, {Rambaux}, {Ramos}, {Ramos-Lerate}, {Re Fiorentin}, {Regibo}, {Reyl{\'e}}, {Ripepi}, {Riva}, {Rixon}, {Robichon}, {Robin}, {Roelens}, {Rohrbasser}, {Romero-G{\'o}mez}, {Rowell}, {Royer}, {Rybicki}, {Sadowski}, {Sagrist{\`a} Sell{\'e}s}, {Sahlmann}, {Salgado}, {Salguero}, {Samaras}, {Sanchez Gimenez}, {Sanna}, {Santove{\~n}a}, {Sarasso}, {Schultheis}, {Sciacca}, {Segol}, {Segovia}, {S{\'e}gransan}, {Semeux}, {Shahaf}, {Siddiqui}, {Siebert}, {Siltala}, {Slezak}, {Smart}, {Solano}, {Solitro}, {Souami}, {Souchay}, {Spagna}, {Spoto}, {Steele}, {Steidelm{\"u}ller}, {Stephenson}, {S{\"u}veges}, {Szabados}, {Szegedi-Elek}, {Taris}, {Tauran}, {Taylor}, {Teixeira}, {Thuillot}, {Tonello}, {Torra}, {Torra}, {Turon}, {Unger}, {Vaillant}, {van Dillen}, {Vanel}, {Vecchiato}, {Viala}, {Vicente},
  {Voutsinas}, {Weiler}, {Wevers}, {Wyrzykowski}, {Yoldas}, {Yvard}, {Zhao}, {Zorec}, {Zucker}, {Zurbach}, \& {Zwitter}}]{Gaia21}
{Gaia Collaboration}, {Brown}, A.~G.~A., {Vallenari}, A., {et~al.} 2021, \aap, 649, A1, \dodoi{10.1051/0004-6361/202039657}

\bibitem[{{Gullberg} {et~al.}(2015){Gullberg}, {De Breuck}, {Vieira}, {Wei{\ss}}, {Aguirre}, {Aravena}, {B{\'e}thermin}, {Bradford}, {Bothwell}, {Carlstrom}, {Chapman}, {Fassnacht}, {Gonzalez}, {Greve}, {Hezaveh}, {Holzapfel}, {Husband}, {Ma}, {Malkan}, {Marrone}, {Menten}, {Murphy}, {Reichardt}, {Spilker}, {Stark}, {Strandet}, \& {Welikala}}]{Gullberg15}
{Gullberg}, B., {De Breuck}, C., {Vieira}, J.~D., {et~al.} 2015, \mnras, 449, 2883, \dodoi{10.1093/mnras/stv372}

\bibitem[{{Hashimoto} {et~al.}(2019){Hashimoto}, {Inoue}, {Mawatari}, {Tamura}, {Matsuo}, {Furusawa}, {Harikane}, {Shibuya}, {Knudsen}, {Kohno}, {Ono}, {Zackrisson}, {Okamoto}, {Kashikawa}, {Oesch}, {Ouchi}, {Ota}, {Shimizu}, {Taniguchi}, {Umehata}, \& {Watson}}]{Hashimoto19}
{Hashimoto}, T., {Inoue}, A.~K., {Mawatari}, K., {et~al.} 2019, \pasj, 71, 71, \dodoi{10.1093/pasj/psz049}

\bibitem[{{Kaufman} {et~al.}(2006){Kaufman}, {Wolfire}, \& {Hollenbach}}]{Kaufman06}
{Kaufman}, M.~J., {Wolfire}, M.~G., \& {Hollenbach}, D.~J. 2006, \apj, 644, 283, \dodoi{10.1086/503596}

\bibitem[{{Klamer} {et~al.}(2004){Klamer}, {Ekers}, {Sadler}, \& {Hunstead}}]{Klamer04}
{Klamer}, I.~J., {Ekers}, R.~D., {Sadler}, E.~M., \& {Hunstead}, R.~W. 2004, \apjl, 612, L97, \dodoi{10.1086/424843}

\bibitem[{{Klamer} {et~al.}(2005){Klamer}, {Ekers}, {Sadler}, {Weiss}, {Hunstead}, \& {De Breuck}}]{Klamer05}
{Klamer}, I.~J., {Ekers}, R.~D., {Sadler}, E.~M., {et~al.} 2005, \apjl, 621, L1, \dodoi{10.1086/429147}

\bibitem[{{Kolwa} {et~al.}(2019){Kolwa}, {Vernet}, {De Breuck}, {Villar-Mart{\'\i}n}, {Humphrey}, {Arrigoni-Battaia}, {Gullberg}, {Falkendal}, {Drouart}, {Lehnert}, {Wylezalek}, \& {Man}}]{Kolwa19}
{Kolwa}, S., {Vernet}, J., {De Breuck}, C., {et~al.} 2019, \aap, 625, A102, \dodoi{10.1051/0004-6361/201935437}

\bibitem[{{Kurk} {et~al.}(2004){Kurk}, {Pentericci}, {R{\"o}ttgering}, \& {Miley}}]{Kurk04}
{Kurk}, J.~D., {Pentericci}, L., {R{\"o}ttgering}, H.~J.~A., \& {Miley}, G.~K. 2004, \aap, 428, 793, \dodoi{10.1051/0004-6361:20040075}

\bibitem[{{Lee} {et~al.}(2023){Lee}, {Kohno}, {Hatsukade}, {Egusa}, {Yamashita}, {Schramm}, {Ichikawa}, {Imanishi}, {Izumi}, {Nagao}, {Toba}, \& {Umehata}}]{Lee23}
{Lee}, K., {Kohno}, K., {Hatsukade}, B., {et~al.} 2023, \apj, 944, 35, \dodoi{10.3847/1538-4357/acaf58}

\bibitem[{{Madden} {et~al.}(2020){Madden}, {Cormier}, {Hony}, {Lebouteiller}, {Abel}, {Galametz}, {De Looze}, {Chevance}, {Polles}, {Lee}, {Galliano}, {Lambert-Huyghe}, {Hu}, \& {Ramambason}}]{Madden20}
{Madden}, S.~C., {Cormier}, D., {Hony}, S., {et~al.} 2020, \aap, 643, A141, \dodoi{10.1051/0004-6361/202038860}

\bibitem[{{Magliocchetti}(2022)}]{Magliocchetti22}
{Magliocchetti}, M. 2022, \aapr, 30, 6, \dodoi{10.1007/s00159-022-00142-1}

\bibitem[{{Matsuoka} {et~al.}(2011){Matsuoka}, {Nagao}, {Maiolino}, {Marconi}, \& {Taniguchi}}]{Matsuoka11}
{Matsuoka}, K., {Nagao}, T., {Maiolino}, R., {Marconi}, A., \& {Taniguchi}, Y. 2011, \aap, 532, L10, \dodoi{10.1051/0004-6361/201117641}

\bibitem[{{Meynet} \& {Maeder}(2003)}]{Meynet03}
{Meynet}, G., \& {Maeder}, A. 2003, \aap, 404, 975, \dodoi{10.1051/0004-6361:20030512}

\bibitem[{{Miley} \& {De Breuck}(2008)}]{MileyDeBeruck08}
{Miley}, G., \& {De Breuck}, C. 2008, \aapr, 15, 67, \dodoi{10.1007/s00159-007-0008-z}

\bibitem[{{Nakajima} {et~al.}(2018){Nakajima}, {Fletcher}, {Ellis}, {Robertson}, \& {Iwata}}]{Nakajima18}
{Nakajima}, K., {Fletcher}, T., {Ellis}, R.~S., {Robertson}, B.~E., \& {Iwata}, I. 2018, \mnras, 477, 2098, \dodoi{10.1093/mnras/sty750}

\bibitem[{{Overzier} {et~al.}(2006){Overzier}, {Miley}, {Bouwens}, {Cross}, {Zirm}, {Ben{\'\i}tez}, {Blakeslee}, {Clampin}, {Demarco}, {Ford}, {Hartig}, {Illingworth}, {Martel}, {R{\"o}ttgering}, {Venemans}, {Ardila}, {Bartko}, {Bradley}, {Broadhurst}, {Coe}, {Feldman}, {Franx}, {Golimowski}, {Goto}, {Gronwall}, {Holden}, {Homeier}, {Infante}, {Kimble}, {Krist}, {Mei}, {Menanteau}, {Meurer}, {Motta}, {Postman}, {Rosati}, {Sirianni}, {Sparks}, {Tran}, {Tsvetanov}, {White}, \& {Zheng}}]{Overzier06}
{Overzier}, R.~A., {Miley}, G.~K., {Bouwens}, R.~J., {et~al.} 2006, \apj, 637, 58, \dodoi{10.1086/498234}

\bibitem[{{Pentericci} {et~al.}(2000){Pentericci}, {Kurk}, {R{\"o}ttgering}, {Miley}, {van Breugel}, {Carilli}, {Ford}, {Heckman}, {McCarthy}, \& {Moorwood}}]{Pentericci00}
{Pentericci}, L., {Kurk}, J.~D., {R{\"o}ttgering}, H.~J.~A., {et~al.} 2000, \aap, 361, L25, \dodoi{10.48550/arXiv.astro-ph/0008143}

\bibitem[{{Pound} \& {Wolfire}(2008)}]{Pound08}
{Pound}, M.~W., \& {Wolfire}, M.~G. 2008, in Astronomical Society of the Pacific Conference Series, Vol. 394, Astronomical Data Analysis Software and Systems XVII, ed. R.~W. {Argyle}, P.~S. {Bunclark}, \& J.~R. {Lewis}, 654

\bibitem[{{Pound} \& {Wolfire}(2011)}]{Pound11}
{Pound}, M.~W., \& {Wolfire}, M.~G. 2011, {PDRT: Photo Dissociation Region Toolbox}.
\newblock \doeprint{1102.022}

\bibitem[{{Pound} \& {Wolfire}(2023)}]{Pound23}
---. 2023, \aj, 165, 25, \dodoi{10.3847/1538-3881/ac9b1f}

\bibitem[{{Reuland} {et~al.}(2004){Reuland}, {R{\"o}ttgering}, {van Breugel}, \& {De Breuck}}]{Reuland04}
{Reuland}, M., {R{\"o}ttgering}, H., {van Breugel}, W., \& {De Breuck}, C. 2004, \mnras, 353, 377, \dodoi{10.1111/j.1365-2966.2004.08063.x}

\bibitem[{{Rose} {et~al.}(2018){Rose}, {Tadhunter}, {Ramos Almeida}, {Rodr{\'\i}guez Zaur{\'\i}n}, {Santoro}, \& {Spence}}]{Rose18}
{Rose}, M., {Tadhunter}, C., {Ramos Almeida}, C., {et~al.} 2018, \mnras, 474, 128, \dodoi{10.1093/mnras/stx2590}

\bibitem[{{Shen} {et~al.}(2020){Shen}, {Hopkins}, {Faucher-Gigu{\`e}re}, {Alexander}, {Richards}, {Ross}, \& {Hickox}}]{Shen20}
{Shen}, X., {Hopkins}, P.~F., {Faucher-Gigu{\`e}re}, C.-A., {et~al.} 2020, \mnras, 495, 3252, \dodoi{10.1093/mnras/staa1381}

\bibitem[{{Shi} {et~al.}(2014){Shi}, {Rieke}, {Ogle}, {Su}, \& {Balog}}]{Shi14}
{Shi}, Y., {Rieke}, G.~H., {Ogle}, P.~M., {Su}, K.~Y.~L., \& {Balog}, Z. 2014, \apjs, 214, 23, \dodoi{10.1088/0067-0049/214/2/23}

\bibitem[{{Smol{\v{c}}i{\'c}} {et~al.}(2017){Smol{\v{c}}i{\'c}}, {Novak}, {Delvecchio}, {Ceraj}, {Bondi}, {Delhaize}, {Marchesi}, {Murphy}, {Schinnerer}, {Vardoulaki}, \& {Zamorani}}]{Smolcic17}
{Smol{\v{c}}i{\'c}}, V., {Novak}, M., {Delvecchio}, I., {et~al.} 2017, \aap, 602, A6, \dodoi{10.1051/0004-6361/201730685}

\bibitem[{{Tang} {et~al.}(2023){Tang}, {Stark}, {Chen}, {Mason}, {Topping}, {Endsley}, {Senchyna}, {Plat}, {Lu}, {Whitler}, {Robertson}, \& {Charlot}}]{Tang23}
{Tang}, M., {Stark}, D.~P., {Chen}, Z., {et~al.} 2023, \mnras, 526, 1657, \dodoi{10.1093/mnras/stad2763}

\bibitem[{{U} {et~al.}(2012){U}, {Sanders}, {Mazzarella}, {Evans}, {Howell}, {Surace}, {Armus}, {Iwasawa}, {Kim}, {Casey}, {Vavilkin}, {Dufault}, {Larson}, {Barnes}, {Chan}, {Frayer}, {Haan}, {Inami}, {Ishida}, {Kartaltepe}, {Melbourne}, \& {Petric}}]{U12}
{U}, V., {Sanders}, D.~B., {Mazzarella}, J.~M., {et~al.} 2012, \apjs, 203, 9, \dodoi{10.1088/0067-0049/203/1/9}

\bibitem[{{van Breugel} {et~al.}(1999){van Breugel}, {De Breuck}, {Stanford}, {Stern}, {R{\"o}ttgering}, \& {Miley}}]{vanBreugel99}
{van Breugel}, W., {De Breuck}, C., {Stanford}, S.~A., {et~al.} 1999, \apjl, 518, L61, \dodoi{10.1086/312080}

\bibitem[{{Veilleux} {et~al.}(2020){Veilleux}, {Maiolino}, {Bolatto}, \& {Aalto}}]{Veilleux20}
{Veilleux}, S., {Maiolino}, R., {Bolatto}, A.~D., \& {Aalto}, S. 2020, \aapr, 28, 2, \dodoi{10.1007/s00159-019-0121-9}

\bibitem[{{Venemans} {et~al.}(2002){Venemans}, {Kurk}, {Miley}, {R{\"o}ttgering}, {van Breugel}, {Carilli}, {De Breuck}, {Ford}, {Heckman}, {McCarthy}, \& {Pentericci}}]{Venemans02}
{Venemans}, B.~P., {Kurk}, J.~D., {Miley}, G.~K., {et~al.} 2002, \apjl, 569, L11, \dodoi{10.1086/340563}

\bibitem[{{Venemans} {et~al.}(2004){Venemans}, {R{\"o}ttgering}, {Overzier}, {Miley}, {De Breuck}, {Kurk}, {van Breugel}, {Carilli}, {Ford}, {Heckman}, {McCarthy}, \& {Pentericci}}]{Venemans04}
{Venemans}, B.~P., {R{\"o}ttgering}, H.~J.~A., {Overzier}, R.~A., {et~al.} 2004, \aap, 424, L17, \dodoi{10.1051/0004-6361:200400041}

\bibitem[{{Verhamme} {et~al.}(2015){Verhamme}, {Orlitov{\'a}}, {Schaerer}, \& {Hayes}}]{Verhamme15}
{Verhamme}, A., {Orlitov{\'a}}, I., {Schaerer}, D., \& {Hayes}, M. 2015, \aap, 578, A7, \dodoi{10.1051/0004-6361/201423978}

\bibitem[{{Verhamme} {et~al.}(2006){Verhamme}, {Schaerer}, \& {Maselli}}]{Verhamme06}
{Verhamme}, A., {Schaerer}, D., \& {Maselli}, A. 2006, \aap, 460, 397, \dodoi{10.1051/0004-6361:20065554}

\bibitem[{{Villar-Mart{\'\i}n} {et~al.}(2003){Villar-Mart{\'\i}n}, {Vernet}, {di Serego Alighieri}, {Fosbury}, {Humphrey}, \& {Pentericci}}]{Villar-Martin03}
{Villar-Mart{\'\i}n}, M., {Vernet}, J., {di Serego Alighieri}, S., {et~al.} 2003, \mnras, 346, 273, \dodoi{10.1046/j.1365-2966.2003.07090.x}

\bibitem[{Virtanen {et~al.}(2020)Virtanen, Gommers, Oliphant, Haberland, Reddy, Cournapeau, Burovski, Peterson, Weckesser, Bright, {van der Walt}, Brett, Wilson, Millman, Mayorov, Nelson, Jones, Kern, Larson, Carey, Polat, Feng, Moore, {VanderPlas}, Laxalde, Perktold, Cimrman, Henriksen, Quintero, Harris, Archibald, Ribeiro, Pedregosa, {van Mulbregt}, \& {SciPy 1.0 Contributors}}]{SciPy20}
Virtanen, P., Gommers, R., Oliphant, T.~E., {et~al.} 2020, Nature Methods, 17, 261, \dodoi{10.1038/s41592-019-0686-2}

\bibitem[{{Wang} {et~al.}(2023){Wang}, {Wylezalek}, {Vernet}, {De Breuck}, {Gullberg}, {Swinbank}, {Villar Mart{\'\i}n}, {Lehnert}, {Drouart}, {Arrigoni Battaia}, {Humphrey}, {Noirot}, {Kolwa}, {Seymour}, \& {Lagos}}]{Wang23}
{Wang}, W., {Wylezalek}, D., {Vernet}, J., {et~al.} 2023, \aap, 680, A70, \dodoi{10.1051/0004-6361/202346415}

\bibitem[{{Yamaguchi} {et~al.}(2019){Yamaguchi}, {Kohno}, {Hatsukade}, {Wang}, {Yoshimura}, {Ao}, {Caputi}, {Dunlop}, {Egami}, {Espada}, {Fujimoto}, {Hayatsu}, {Ivison}, {Kodama}, {Kusakabe}, {Nagao}, {Ouchi}, {Rujopakarn}, {Tadaki}, {Tamura}, {Ueda}, {Umehata}, {Wang}, \& {Yun}}]{Yamaguchi19}
{Yamaguchi}, Y., {Kohno}, K., {Hatsukade}, B., {et~al.} 2019, \apj, 878, 73, \dodoi{10.3847/1538-4357/ab0d22}

\end{thebibliography}
\bibliographystyle{aasjournal}



\end{document}